\documentclass[twocolumn]{aastex631}

\usepackage{amsmath}
\usepackage{amssymb}
\usepackage{booktabs}
\usepackage{graphicx}

\newcommand{\hmpc}{h^{-1}\,\mathrm{Mpc}}
\newcommand{\kms}{\mathrm{km\,s^{-1}}}
\newcommand{\dmu}{\Delta\mu}
\newcommand{\Vdip}{\boldsymbol{V}_{\mathrm{dip}}}
\newcommand{\lcdm}{\ensuremath{\Lambda\mathrm{CDM}}}

\graphicspath{{figures/}}

\shorttitle{Large-Scale Motions in Cosmicflows-4}
\shortauthors{Nusser \& Tully}

\begin{document}

\title{Cosmicflows-4 and the Cosmological Consistency of Large-Scale
Motions}

\author[0000-0002-8272-4779]{Adi Nusser}
\email{adin@technion.ac.il}
\affiliation{Department of Physics, Technion -- Israel Institute of
Technology, Haifa 3200003, Israel}

\author{R. Brent Tully}
\email{tully@ifa.hawaii.edu}
\affiliation{Institute for Astronomy, University of Hawaii at M\={a}noa,
2680 Woodlawn Drive, Honolulu, HI 96822, USA}

\begin{abstract}
We study coherent flows in Cosmicflows-4 and constituent catalogs. In
nonoverlapping radial bins, we fit monopole and dipole moments using a
log-distance radial-velocity estimator rather than an ideal-sphere bulk
flow. For each sample, the Planck 2018 \lcdm\ prediction uses the actual
positions and weights, including linear-theory velocity and measurement
covariance. CMB-frame redshift-space binning reduces sensitivity to
distance-placement Malmquist effects. Across six bins through
$300\,\hmpc$, All Individual is broadly consistent with \lcdm: its dipole and joint
deviations are $2.29\sigma$ and $2.36\sigma$. Four bins through
$160\,\hmpc$ give similar values of $2.34\sigma$ and $2.42\sigma$.
Nevertheless, $|\Vdip|$ rises over
$R_{\rm eff}\simeq100$--$150\,\hmpc$. In the
$120$--$160\,\hmpc$ bin it gives $|\Vdip|=628\pm82\,\kms$ and a
dipole deviation of $3.40\sigma$. The interpretation is strongly
window-dependent because component samples differ in radial selection,
angular coverage, and distance precision.
The excess is concentrated toward negative supergalactic X (SGX). TFR and 6dFGS
favor larger dipoles, whereas SDSS and SN favor smaller ones. Removing
6dFGS reduces the upper-bin dipole deviation to $1.56\sigma$, while
removing SN raises it to $3.77\sigma$. Nevertheless, the SN and 6dFGS
dipoles are mutually consistent in a covariance-aware comparison.
SDSS is consistent with \lcdm\ through $160\,\hmpc$ but shows a separate
outer-bin rise whose significance depends on monopole fitting and
a shared distance-scale covariance. Significances use the baseline
covariance model, without cross-catalog calibration covariance or an
adjustment for multiple catalogs, bins, and components.
\end{abstract}


\section{Introduction}
\label{sec:introduction}

In a gravitational theory that obeys the equivalence principle, freely
falling bodies at the same location experience the same gravitational
acceleration, independently of their composition. Peculiar velocities,
defined as deviations from a pure Hubble flow, therefore offer a direct
tracer of the gravitational field. On scales large compared with the sizes
of galaxies, groups, and clusters, their coherent motions are expected to
approach the velocity field of the total matter.
This expectation need not hold object by object or on small scales, where
different assembly histories, smoothing associated with finite halo
masses, nonlinear orbital motions, and hydrodynamical or environmental
forces can produce velocity differences
\citep{SummersDavisEvrard1995,Schmidt2010VelocityBias}.

This approximate universality of coherent large-scale motion contrasts
with the galaxy number-density field. The spatial distribution of a
selected galaxy population is generally biased relative to the underlying
matter density, with clustering that depends, for example, on luminosity
and color \citep{Zehavi2011Clustering}. Its coherent motion, by contrast,
is driven by the gravitational field of the total matter. Just as
rotational speeds in galaxies and internal velocity dispersions in
galaxies and clusters constrain their gravitating mass, coherent galaxy
motions across large volumes probe the cosmological distribution of dark
matter and baryonic matter. In gravitational-instability theory, the
velocity field is related to the mass-density field and its growth.
Peculiar-velocity measurements can therefore test both the distribution
of matter and the growth of structure
\citep{Gorski1989,Nusser2011,Nusser2017VelocityDensity}.

Peculiar-velocity catalogs generally provide only the radial component of
the peculiar velocity. It is estimated by comparing the observed redshift
with the expansion velocity inferred from an independent distance
indicator. Transverse velocities additionally require accurate proper
motions and are available mainly for nearby Local Group systems, through
maser, Hubble Space Telescope, and Gaia measurements
\citep{Brunthaler2005M33,Sohn2012M31,Kallivayalil2013MC,
Battaglia2022Gaia}. Full three-dimensional velocities remain the exception
rather than the basis of cosmological peculiar-velocity catalogs.

Extracting cosmological information from the radial velocities is
nevertheless fraught with difficulties. Intrinsic scatter and
heterogeneity among distance indicators, sample selection, calibration
and zero-point errors, sparse sampling, and incomplete angular coverage
can all produce systematic effects. In particular, noisy distances lead
to distance Malmquist bias, while placing
objects at those noisy distances can generate velocity Malmquist bias and
spurious velocity gradients. These effects are distinct from the skewness
introduced when approximately Gaussian distance-modulus errors are
converted nonlinearly into distances and velocities
\citep{LyndenBell1988,StraussWillick1995,NusserDavis1995,Hoffman2021,
Nusser2026DistanceVelocity}. Placing galaxies at their redshift-space
positions reduces sensitivity to these Malmquist effects, although it does
not remove calibration, selection, or angular-window uncertainties
\citep{Nusser2016BulkFlows,Nusser2026DistanceVelocity}.

Despite these difficulties, comparisons of observed peculiar velocities
with the gravitational flows predicted from galaxy redshift surveys have
generally found broad agreement on large scales. Such agreement has been
reported for SFI++ velocities compared with 2MRS, Tully--Fisher and
supernova velocities compared with 2M++, and Cosmicflows-3 velocities
compared with 2MRS
\citep{Davis2011LocalGravity,Carrick2015,Nusser2017VelocityDensity,
LilowNusser2021}. A recent field-level Bayesian reconstruction, with
posterior sampling by Hamiltonian Monte Carlo, likewise finds good
agreement between the gravitational flow reconstructed from 2MRS and the
observed CF4 group velocities \citep{Nusser2026Bayesian2MRS}.

Against this background, and separately from comparisons with density
fields reconstructed from redshift surveys, interest has focused on
whether Cosmicflows-4 (CF4) contains unusually large coherent motions on
the largest scales probed by the catalog. CF4 combines several distance
indicators in two compilations of the Extragalactic Distance Database
(EDD; \citealt{Tully2009EDD}).\footnote{https://edd.ifa.\allowbreak hawaii.edu}
The compilations are ``All CF4 Individual Distances'' and ``CF4 All
Groups,'' which we call All Individual and
Grouped, respectively \citep{CF4}. Applying the same
fits to both compilations gives nearly coincident best-fitting moments.
Because they share measurements and do not provide independent tests, we
show results only for All Individual below. Coherent motions on these
scales are usually quantified by the bulk flow, defined as the
window-weighted mean three-dimensional peculiar-velocity vector within a
specified sphere centered on the observer. Both its estimated amplitude
and its predicted \lcdm\ variance depend on the radial and angular survey
window \citep{feldwh10,Nusser2014a,Nusser2016BulkFlows}.

Several analyses of CF4 have reported a large bulk flow on scales of
roughly $120$--$200\,\hmpc$, although the estimated significance depends
on the method. \citet{Watkins2023CF4BulkFlow} obtained
$395\pm29\,\kms$ for a minimum-variance estimate targeting a sphere of
radius $150\,\hmpc$, corresponding to $3.8\sigma$ in two-sided Gaussian
notation. \citet{Whitford2023CF4Estimators} found
$428\pm108\,\kms$ at an effective depth of $173\,\hmpc$, corresponding
to $3.3\sigma$. They also showed that survey geometry and underestimated
estimator uncertainties can make the disagreement with \lcdm\ appear
stronger. In a Wiener-filter reconstruction,
\citet{Hoffman2024CF4Velocity} found that 6dFGS dominates the bulk-flow
profile beyond approximately $120\,\hmpc$. It drives an excess that
reaches about $3.4\sigma$ at $250\,\hmpc$, mainly through the SGX
component in the approximate direction of the Shapley Concentration.
A prior-free CF4 analysis found bulk-flow tensions near $140$ and
$240\,\hmpc$ \citep{Duangchan2026CF4}.

Other analyses give a more cautious interpretation. The CF4++ velocity
field gives $315\pm40\,\kms$ inside $150\,\hmpc$ and describes the
average disagreement over $100$--$150\,\hmpc$ as approximately
$1.7\sigma$ \citep{Courtois2025CF4pp}. It also warns that anisotropic
coverage affects the result at greater distances. Tests of the 2M++
velocity--density comparison likewise find that previously reported
residual bulk flows can be consistent with \lcdm\ after reconstruction
and selection effects are included \citep{Hollinger2024TwoMpp}.
These results are not direct measurements of the same quantity.
They use cumulative top-hat spheres, minimum-variance windows, or
reconstructed three-dimensional fields.

Previous CF4 bulk-flow studies have generally emphasized the combined All
Individual or Grouped compilations rather than systematically resolving the
contributions of their constituent catalogs. This approach
can obscure the heterogeneous structure of CF4. The component samples
differ in distance precision, calibration, radial selection, and angular
coverage. With uneven sky coverage, a change in the monopole can be partly
offset by a change in the dipole. A large combined-catalog flow therefore
does not by itself establish whether the signal is shared among the
components or driven primarily by a particular sample.

We address this limitation through a catalog-by-catalog analysis in
the CMB redshift frame. We test both the combined All Individual catalog
and its constituent samples for consistency with \lcdm, and identify which
components drive any excess large-scale motion. We do not attempt to
reconstruct the volume-averaged velocity within an ideal, fully sampled
sphere, since that would require interpolating the sparse and anisotropic
measurements throughout the survey volume. Instead, we express the
distance differences as linear log-distance velocities and fit
monopole-only, dipole-only, and joint models in nonoverlapping radial bins.
For every sample and bin, we evaluate the \lcdm\ velocity covariance at
the actual galaxy positions and use it, together with the measurement
covariance, in the same weighted estimator. The
resulting prediction accounts for the radial and angular galaxy
distribution and the statistical weights, rather than assuming an
idealized spherical survey window. This approach also shows directly how
the sky coverage couples the fitted monopole to the dipole components.

Section~\ref{sec:data} describes the samples and frame.
Section~\ref{sec:estimator} defines the distance conversion and fits.
Section~\ref{sec:covariance} gives the measurement and \lcdm\ covariances.
Section~\ref{sec:results} presents the results.
Section~\ref{sec:discussion} discusses robustness, limitations, and
interpretation. Section~\ref{sec:conclusions} summarizes the main findings.

\section{Data and coordinate convention}
\label{sec:data}

\subsection{Catalog samples}

Our combined-catalog results use the CF4 All Individual parent catalog.
The main figures show four component catalogs. SDSS is the Sloan
Digital Sky Survey Fundamental Plane (FP) sample. 6dFGS is the 6dF Galaxy
Survey FP sample. TFR is the Tully--Fisher relation sample.
SN combines the valid SN Ia and SN II distances. If both estimates are
available for one object, we combine them with inverse-variance weights.
For SN II, CF4 standardizes the luminosity using the photospheric
expansion velocity and color, giving individual distances with roughly
15 percent scatter \citep{CF4}.
Other FP combines the smaller Fundamental Plane data sets labeled SMAC,
ENEAR, and EFAR in CF4. It remains in the numerical tables and statistical
tests but is omitted from the figures to reduce crowding. We also retain the
tip of the red giant branch sample, denoted TRGB, in the catalog accounting.
The SDSS peculiar-velocity
construction is described by
\citet{Howlett2022SDSSPV}. Relevant 6dFGS and TFR catalog descriptions are
given by \citet{Springob2014} and \citet{Kourkchi2020}.
The smaller SBF, TRGB, Cepheid, and maser contributions remain part of
All Individual but are not fitted as separate hard-bin curves.

Table~\ref{tab:catalog_composition} combines the full catalog totals with
their radial composition. TRGB is retained as its own distance indicator,
but it does not meet the support requirements
in any of the hard bins and is absent from the contour and probability
tables. Catalog counts are not additive because a galaxy can have more than
one distance estimate.

\begin{table}[t]
\centering
\scriptsize
\setlength{\tabcolsep}{2.2pt}
\caption{Catalog composition and radial support}
\label{tab:catalog_composition}
\begin{tabular}{@{}lrrrrrr@{}}
\toprule
Bin & \multicolumn{1}{c}{All} & SDSS & 6dFGS & TFR & SN &
\multicolumn{1}{c}{Other} \\
($\hmpc$) & \multicolumn{1}{c}{Individual} & & & & &
\multicolumn{1}{c}{FP} \\
\midrule
$0$--$40$    &  4,227 &    44 &   172 & 3,273 & 182 & 155 \\
$40$--$60$   &  4,239 &   210 &   559 & 3,180 & 176 & 230 \\
$60$--$90$   &  6,585 & 1,567 & 1,156 & 3,451 & 238 & 371 \\
$90$--$120$  &  6,767 & 2,583 & 2,097 & 1,654 & 178 & 440 \\
$120$--$160$ &  8,481 & 4,606 & 3,097 &   501 & 108 & 254 \\
$160$--$200$ &  6,347 & 6,099 &    16 &   111 &  93 &  36 \\
$200$--$300$ & 19,101 &18,934 &     0 &    42 & 116 &  16 \\
\addlinespace
All radii     & 55,749 &34,043 & 7,097 &12,212 &1,093&1,502 \\
\bottomrule
\end{tabular}
\tablecomments{Catalog columns can overlap and must not be summed. The
``All radii'' row gives the full catalog totals before the hard-bin cuts.
Other FP combines the SMAC, ENEAR, and EFAR FP samples.}
\end{table}

\begin{table}[t]
\centering
\scriptsize
\setlength{\tabcolsep}{3.0pt}
\caption{Direct overlap between component catalogs}
\label{tab:catalog_overlap}
\textbf{(a) Selected full-sample overlaps}\\[3pt]
\begin{tabular}{@{}llrr@{}}
\toprule
Catalog & Catalog & Shared & Shared fraction \\
A & B & objects & of smaller catalog \\
\midrule
SDSS    & Other FP       &    251 & 0.167 \\
6dFGS   & Other FP       &    133 & 0.0885 \\
6dFGS   & TFR            &     60 & 0.00845 \\
SDSS    & 6dFGS          &     41 & 0.00578 \\
SDSS    & TFR            &      2 & $1.64\times10^{-4}$ \\
\bottomrule
\end{tabular}

\vspace{6pt}
\textbf{(b) Overlap by bin}\\[3pt]
\setlength{\tabcolsep}{1.6pt}
\begin{tabular}{@{}lrrrrrr@{}}
\toprule
Bin & SDSS-- & SDSS-- & SDSS-- & 6dFGS-- & 6dFGS-- & TFR-- \\
($\hmpc$) & 6dFGS & TFR & Other FP & TFR & Other FP & Other FP \\
\midrule
$40$--$60$   &  1 & 0 & 10 & 14 & 38 & 1 \\
$60$--$90$   &  6 & 1 & 97 & 23 & 19 & 1 \\
$90$--$120$  & 13 & 0 & 97 & 12 & 38 & 2 \\
$120$--$160$ & 21 & 0 & 33 &  2 & 24 & 1 \\
$160$--$200$ &  0 & 0 &  5 &  0 &  0 & 0 \\
\bottomrule
\end{tabular}
\tablecomments{
Panel (a) gives the shared fraction of the smaller catalog. Panel (b) counts objects in the same radial bin in
both catalogs, with the radius evaluated separately for each catalog.}
\end{table}

Table~\ref{tab:catalog_overlap} shows that direct overlap between
different component catalogs is small in every bin. These counts describe
radial overlap only. They do not show whether the catalogs cover the same
directions on the sky.

\subsection{Frame, positions, and radial bins}

All redshifts used in this paper are in the CMB frame. This frame removes
the observer velocity inferred from the cosmic microwave background
dipole. To reduce sensitivity to Malmquist effects associated with noisy
distance placement
\citep{StraussWillick1995,Nusser2016BulkFlows,
Nusser2026DistanceVelocity}, we use the
CMB-frame redshift, rather than the
measured distance modulus, to assign both the position and radial bin. An object is
therefore placed at
\begin{equation}
 \boldsymbol{x}_i
 =
 hD_C(z_{{\rm CMB},i})\,
 \widehat{\boldsymbol{r}}_i,
 \label{eq:redshift_position}
\end{equation}
where $z_{{\rm CMB},i}$ is its CMB-frame redshift, $D_C(z)$ is the
comoving distance derived in the \lcdm\ background cosmology, and
$\widehat{\boldsymbol{r}}_i$ is the unit vector from the observer to the
object. We adopt a spatially flat Planck 2018 \lcdm\ cosmology with
$H_0=67.66\,\kms\,\mathrm{Mpc}^{-1}$,
$\Omega_{{\rm m},0}=0.3111$, $\sigma_8=0.8102$, and
$n_s=0.9665$ \citep{PlanckCollaboration2018}. We write
$r_i=|\boldsymbol{x}_i|=hD_C(z_{{\rm CMB},i})$, where
$h=H_0/(100\,\kms\,\mathrm{Mpc}^{-1})$. Distances quoted in
$h^{-1}\,\mathrm{Mpc}$ therefore do not depend on the numerical choice of
$H_0$. The direction vector and dipole are expressed in supergalactic
Cartesian coordinates, whose axes are denoted SGX, SGY, and SGZ.

The radial bins used in the reported fits are listed in
Table~\ref{tab:joint_results}. A bin is retained when it contains at
least 10 catalog objects and satisfies
\begin{equation}
 N_{\rm eff}
 =
 \frac{\left(\sum_i w_i\right)^2}{\sum_i w_i^2}
 \geq 10,
 \label{eq:neff}
\end{equation}
where $w_i$ is the inverse measurement-variance weight used in the velocity
fit and defined in Equation~\eqref{eq:sigmau}. The quantity $N_{\rm eff}$
is the effective number of equally weighted objects. It becomes smaller
than the actual object count when a small number of objects carry most of
the weight. Requiring $N_{\rm eff}\geq10$ does not guarantee uniform sky
coverage. We examine the sky directions and fit covariances separately.

These criteria determine where individual catalog curves are shown. The
noisy 6dFGS $160$--$200\,\hmpc$ bin is omitted. Other FP is retained in
the calculation but is not plotted. SDSS, SN, and All Individual are
also evaluated in the outermost $200$--$300\,\hmpc$ bin. The SN sample
contains 116 objects there, whereas SDSS dominates the separately plotted
component catalogs. These outer estimates should be interpreted
using the quoted covariance rather than their amplitudes alone.

\begin{figure*}[t]
\centering
\includegraphics[width=0.92\textwidth]
{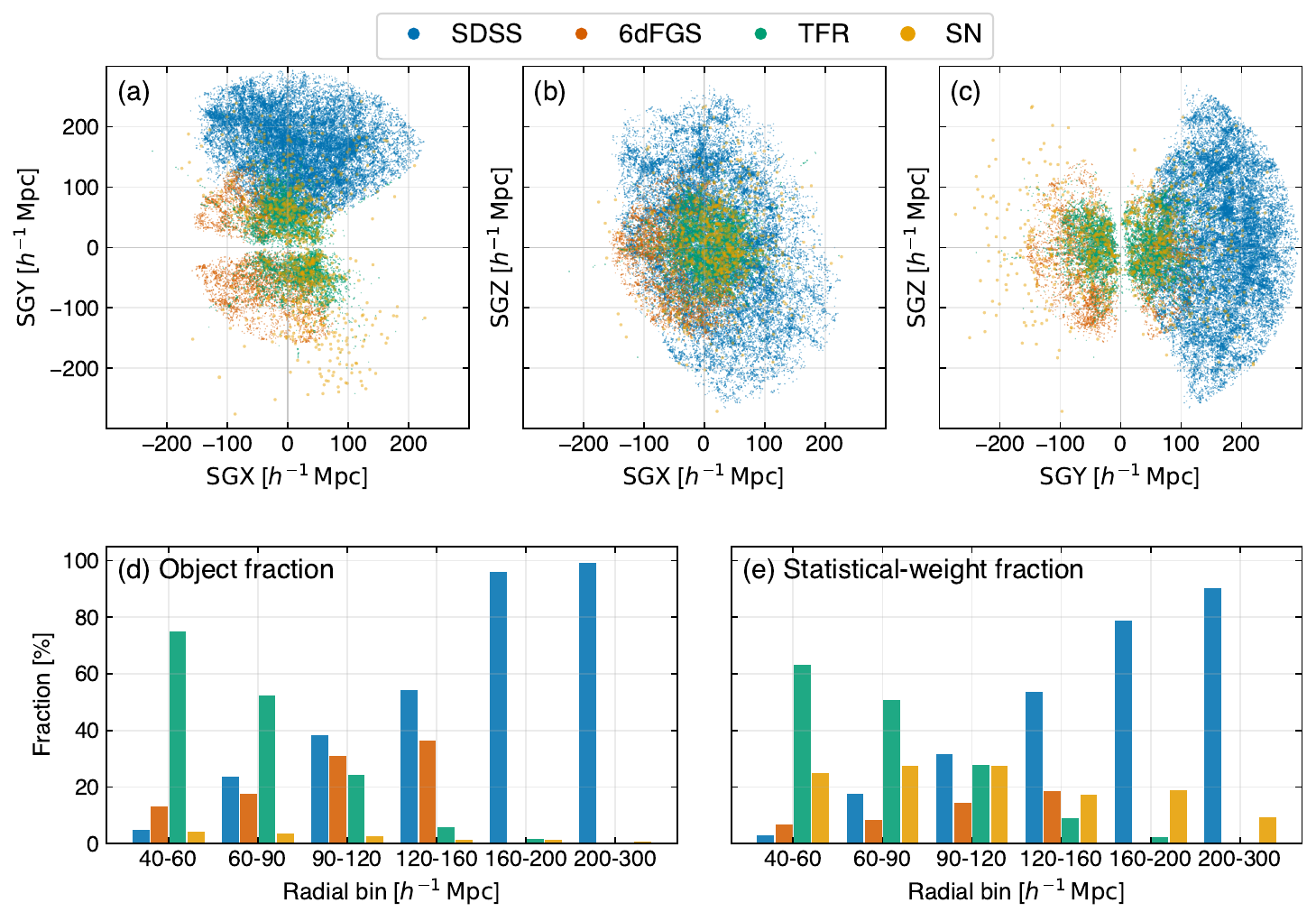}
\caption{Angular and radial composition of the main component catalogs in
All Individual. The top row shows accepted objects at
$40\leq r<300\,\hmpc$. Panels (d) and (e) show each catalog's fraction
of the objects and inverse-variance fit weight in each bin.}
\label{fig:catalog_angular_overlap}
\end{figure*}

The top row of Figure~\ref{fig:catalog_angular_overlap} shows the angular
and radial coverage of the four main component catalogs. SDSS lies mainly
at positive SGY and extends to the largest distances. The 6dFGS sample is
concentrated toward negative SGX, while SN is sparse and broadly
distributed. The $60$--$120\,\hmpc$ range provides the most useful direct
comparison of SDSS, 6dFGS, and TFR because all three have substantial
support there. TFR declines rapidly beyond about $100\,\hmpc$, with 501
objects in the $120$--$160\,\hmpc$ bin and 111 in the
$160$--$200\,\hmpc$ bin, compared with 6,099 SDSS objects in the latter
bin. Beyond $160\,\hmpc$, SDSS is the only major component catalog with
substantial statistical support.

Panels (d) and (e) compare the fraction of catalog objects
with the fraction of the total fit weight in each radial bin. SN is much
more prominent in panel (e) because its smaller distance-modulus errors
give it larger inverse-variance weights. Object fraction is therefore not
a direct measure of a catalog's leverage. Table~\ref{tab:catalog_composition}
gives the corresponding object counts.

For a more quantitative description of the angular coverage, we use the
weighted mean
\begin{equation}
 \langle q\rangle_w
 =
 \frac{\sum_i w_i q_i}{\sum_i w_i},
 \qquad
 w_i=\sigma_{u,i}^{-2},
 \label{eq:weighted_mean}
\end{equation}
where $\sigma_{u,i}$ is the velocity uncertainty defined in
Equation~\eqref{eq:sigmau}. The horizontal coordinate assigned to each
radial bin is
\begin{equation}
 R_{\rm eff}
 =
 \langle r\rangle_w
 =
 \frac{\sum_i w_i r_i}{\sum_i w_i},
 \label{eq:reff}
\end{equation}
the weighted mean redshift-space radius of the catalog objects in that
bin.

\begin{figure*}[t]
\centering
\includegraphics[width=0.94\textwidth]
{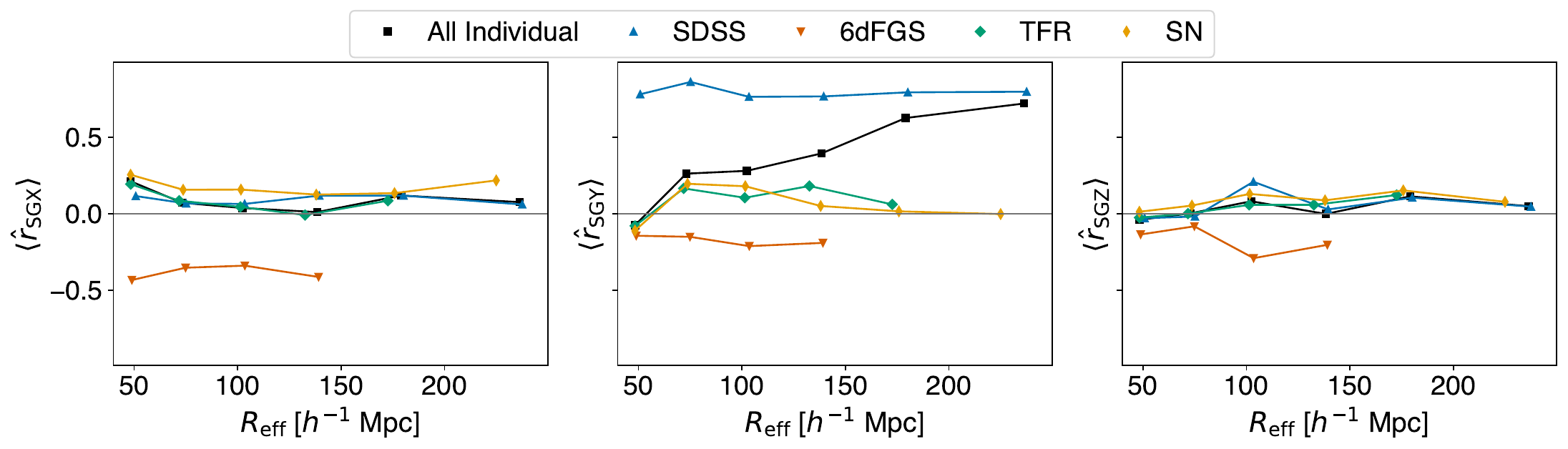}
\caption{Weighted mean supergalactic sightline direction as a function of
$R_{\rm eff}$. A nonzero value shows a preferred survey direction and
couples the monopole to the corresponding dipole component. Curves show
All Individual, SDSS, 6dFGS, TFR, and SN; Other FP is omitted for clarity.}
\label{fig:orientation}
\end{figure*}

Figure~\ref{fig:orientation} shows the three components of the weighted
mean sightline direction as functions of $R_{\rm eff}$. It is therefore a
quantitative summary of the angular coverage entering each fit. A catalog
with balanced sky coverage has all three means near zero. SDSS has a large
positive mean SGY direction, whereas 6dFGS has a negative mean SGX
direction. A nonzero mean direction also couples the fitted monopole to
the corresponding dipole component.

The larger number of objects in one component catalog does not by itself
make its curve coincide with All Individual because the plotted means are
inverse-variance weighted.
The enhanced SN weight seen in Figure~\ref{fig:catalog_angular_overlap}(e)
explains the separation between the All Individual and SDSS curves in
the outer SGY panel of Figure~\ref{fig:orientation}. In the
$160$--$200\,\hmpc$ bin,
6,099 of the 6,347 All Individual objects are accepted SDSS objects. They
make up 96 percent of the objects but only 79 percent of the total
inverse-variance weight. The other 248 objects carry 21 percent of the
weight and have a combined weighted SGY mean close to zero. Ninety of
these objects are
supernova distances, comprising 83 SNe Ia and 7 SNe II. Their median
reported distance-modulus uncertainty is 0.13 mag, and they alone carry
18 percent of the total
weight. They therefore pull the All Individual mean from the SDSS value
of 0.79 to 0.63. The same effect is smaller but still visible in the
$200$--$300\,\hmpc$ bin. There, 18,934 of 19,101 objects are SDSS, while
the 167 non-SDSS objects carry 10 percent of the weight. The latter
include 112 SNe Ia and 3 SNe II. Selecting only the SDSS objects inside
All Individual gives weighted SGY means of 0.79 and 0.80 in the two bins,
essentially identical to the corresponding SDSS values.

\section{Distance conversion and fitting method}
\label{sec:estimator}

\subsection{Distance-modulus difference}

For each object, we define
\begin{equation}
 \dmu_i
 =
 {\rm DM}_i-{\rm DM}_{z,i},
 \label{eq:dmu}
\end{equation}
where the distance modulus is
${\rm DM}=5\log_{10}(D_L/\mathrm{Mpc})+25$ and $D_L$ is luminosity
distance. The values ${\rm DM}_i$ and $e{\rm DM}_i$ are the measured
distance modulus and its reported one-standard-deviation error in the
corresponding CF4 file. The quantity ${\rm DM}_{z,i}$ is the distance
modulus predicted from the CMB-frame redshift when peculiar velocity is
zero. A positive $\dmu_i$ means that the measured distance is larger than
the redshift distance. We use the direct label
${\rm DM}-{\rm DM}_z$ in the figures to avoid implying that all selection
or calibration effects have already been removed.

The first-order velocity estimate made from the
logarithmic distance difference is
\begin{equation}
 u_{\ln,i}
 =
 -K_i a\,\dmu_i,
 \label{eq:ulog}
\end{equation}
where $a=\ln 10/5$ and $K_i$ is the redshift-dependent conversion factor derived in
Appendix~\ref{app:dm_velocity}, which also gives the equivalent fit in
distance-modulus space. It approaches $cz_i$ at low redshift.
The expansion is first order in the dimensionless ratio
$|u_i|/K_i$, which approaches $|u_i|/(cz_i)$ at low redshift. All fitted
objects have $cz_{{\rm CMB},i}>4000\,\kms$, substantially larger than
their expected true peculiar velocities. The peculiar velocity is
therefore a perturbation to the redshift-distance scale. 
The subscript $\ln$ indicates that this estimate comes from a logarithmic
distance. At low redshift Equation~\eqref{eq:ulog} is equivalent to
$\ln(cz_i-u_i)\simeq\ln cz_i-u_i/cz_i$. The generalized $K_i$ form is the
form used in the calculation. Log-distance estimators have long been used
in peculiar-velocity analyses \citep{NusserDavis1995,Nusser2011}. They are
useful because distance-modulus errors are closer to Gaussian than errors
in linear peculiar velocity
\citep{DavisScrimgeour2014,Watkins2015,Hoffman2021}.

The adopted (diagonal) peculiar-velocity measurement error is
\begin{equation}
 \sigma_{u,i}^2
 =
 \left(K_i a\,e{\rm DM}_i\right)^2
 +
 \sigma_{\rm floor}^2,
 \qquad
 \sigma_{\rm floor}=100\,\kms.
 \label{eq:sigmau}
\end{equation}
Thus $\sigma_{u,i}$ is the one-standard-deviation error assigned to
object $i$. The $100\,\kms$ floor prevents objects with very small
reported distance errors from receiving unreasonably large weight. It
represents independent small-scale velocity noise and does not model
correlated nonlinear velocities.

\subsection{Monopole and dipole fits}

Within each radial bin, we approximate the angular dependence of the
radial velocity using only the lowest spherical-harmonic orders. The
constant $m$ corresponds to the $\ell=0$ monopole. The direction cosines
$(\widehat r_x,\widehat r_y,\widehat r_z)$, proportional to the real
$\ell=1$ basis $(\sin\theta\cos\phi,\sin\theta\sin\phi,\cos\theta)$,
define the dipole coefficients collected in the vector $\Vdip$. We fit
three versions of this low-order model:
\begin{align}
 u_{\ln,i} &= m,
 && \Vdip=\boldsymbol{0},
 \label{eq:model_m}\\
 u_{\ln,i} &= \Vdip\mathbin{\cdot}
 \widehat{\boldsymbol{r}}_i,
 && m=0,
 \label{eq:model_v}\\
 u_{\ln,i} &= m+\Vdip\mathbin{\cdot}
 \widehat{\boldsymbol{r}}_i,
 \label{eq:model_joint}
\end{align}
where $u_{\ln,i}$ is the comoving-coordinate radial peculiar velocity of
object $i$, with positive values directed away from the observer, and
$\widehat{\boldsymbol r}_i$ points toward the object. Both $m$ and the
three Cartesian components of $\Vdip$ have units of $\kms$. 
corresponding physical radial peculiar velocity is

This is a low-order parametrization of the velocities sampled by a
catalog, not a reconstruction of the full velocity field or a
volume-averaged flow in an ideal sphere. Because the angular masks differ,
as Figure~\ref{fig:orientation} illustrates, dipoles fitted to two component
catalogs need not represent the same windowed quantity. In All Individual,
the component samples instead contribute jointly to one combined fit.
For the \lcdm\ comparison, the velocity covariance is evaluated using
the actual positions, weights, and estimator of each catalog
(Section~\ref{sec:covariance}), so its angular coverage and cosmic variance
are included in the catalog-specific prediction.

Each of Equations~\eqref{eq:model_m}--\eqref{eq:model_joint} can be
written as
$\boldsymbol u_{\ln}=X\boldsymbol\beta+\boldsymbol\epsilon$.
For a bin containing $N_{\rm obj}$ objects, $\boldsymbol u_{\ln}$ is the
$N_{\rm obj}$-element column vector whose $i$th entry is $u_{\ln,i}$.
The rows of the design matrix and the corresponding parameter vectors
are defined as follows. For Equation~\eqref{eq:model_m},
$X_i=(1)$ and $\boldsymbol\beta=(m)$. For
Equation~\eqref{eq:model_v},
$X_i=(\widehat r_{i,x},\widehat r_{i,y},\widehat r_{i,z})$ and
$\boldsymbol\beta=(V_x,V_y,V_z)^{\mathsf T}$. For
Equation~\eqref{eq:model_joint},
$X_i=(1,\widehat r_{i,x},\widehat r_{i,y},\widehat r_{i,z})$ and
$\boldsymbol\beta=(m,V_x,V_y,V_z)^{\mathsf T}$. Thus $X$ has dimensions
$N_{\rm obj}\times n_{\rm par}$, where $n_{\rm par}=1$, 3, or 4 for
the three models.

Using the same object ordering in $\boldsymbol u_{\ln}$, $X$, and
$W=\operatorname{diag}(\sigma_{u,1}^{-2},\ldots,
\sigma_{u,N_{\rm obj}}^{-2})$, the model parameters are obtained by
weighted least squares. The solution is
\begin{align}
 \widehat{\boldsymbol{\beta}}
 &=A\boldsymbol{u}_{\ln},
 \label{eq:beta}\\
 A
 &=\left(X^{\mathsf T}WX\right)^{-1}X^{\mathsf T}W.
 \label{eq:operator}
\end{align}
For each of Equations~\eqref{eq:model_m}--\eqref{eq:model_joint}, $A$ is
computed using that model's design matrix. It has dimensions
$n_{\rm par}\times N_{\rm obj}$ and maps the measured radial velocities
to the corresponding one, three, or four fitted parameters.

Equivalently, the estimator in Equations~\eqref{eq:beta} and
\eqref{eq:operator} is the value of $\boldsymbol\beta$ that minimizes the
measurement-weighted residual statistic
\begin{equation}
 \chi_{\rm GOF}^2(\boldsymbol\beta)
 =
 \sum_i
 \frac{\left[u_{\ln,i}-(X\boldsymbol\beta)_i\right]^2}
 {\sigma_{u,i}^2},
 \label{eq:gof}
\end{equation}
where $N_{\rm obj}$ is the number of fitted objects and
$n_{\rm par}=1,3$, or 4 is the number of fitted parameters for the three
models. In the joint model, $m,V_x,V_y$, and $V_z$ are minimized
simultaneously. The amplitude
$|\Vdip|=(V_x^2+V_y^2+V_z^2)^{1/2}$ is calculated from the fitted
Cartesian components afterward; it is not itself a fitted or minimized
parameter. The minimized value tests how well a constant monopole and
dipole describe the objects in a bin. Its degrees of freedom are
$\nu=N_{\rm obj}-n_{\rm par}$; for the joint fits reported below,
$\nu=N_{\rm obj}-4$. It is not the \lcdm\ test defined below.

\section{Measurement and \texorpdfstring{\lcdm}{LambdaCDM} covariance}
\label{sec:covariance}

Let $\theta_a$ and $\theta_b$ denote any two fitted quantities chosen
from $\widehat m$, $\widehat V_{\rm SGX}$,
$\widehat V_{\rm SGY}$, and $\widehat V_{\rm SGZ}$. Their covariance is
$C_{ab}=\langle(\theta_a-\langle\theta_a\rangle)
(\theta_b-\langle\theta_b\rangle)\rangle$. A diagonal element is a
variance. An off-diagonal element describes how the two quantities vary
together.

\subsection{Measurement covariance}

Let
\begin{equation}
 C_\epsilon
 =
 \operatorname{diag}(\sigma_{u,1}^2,\ldots,\sigma_{u,N_{\rm obj}}^2).
\end{equation}
The covariance of the fitted moments due to measurement errors is
\begin{equation}
 C_{\rm meas}
 =
 A C_\epsilon A^{\mathsf T}.
 \label{eq:cmeas}
\end{equation}
The matrix $C_\epsilon$ contains the assigned object errors. It is
diagonal because this calculation treats different object errors as
independent. Equation~\eqref{eq:cmeas} maps those object errors into the
covariance of the fitted $m$, $V_x$, $V_y$, and $V_z$. It does not include shared distance
zero-point errors or covariance between different catalogs. The dedicated
SDSS analysis in Section~\ref{sec:sdss} uses a shared two-percent
distance-scale uncertainty as an illustrative correlated-covariance test.

\subsection{Linear velocity covariance}

The radial velocity covariance
$S_{ij}=\langle u_i u_j\rangle_{\lcdm}$ is obtained by projecting the
Cartesian peculiar-velocity covariance along the two lines of sight. For
objects $i$ and $j$, let
$\boldsymbol{s}_{ij}=\boldsymbol{x}_i-\boldsymbol{x}_j$. In linear theory,
the projection gives \citep{Gorski1989}
\begin{align}
 S_{ij}
 ={}&
 \Psi_\perp(s_{ij})
 \left(
 \widehat{\boldsymbol{r}}_i\mathbin{\cdot}
 \widehat{\boldsymbol{r}}_j
 \right)
 \nonumber\\
 &+
 \left[
 \Psi_\parallel(s_{ij})-\Psi_\perp(s_{ij})
 \right]
 \left(
 \widehat{\boldsymbol{r}}_i\mathbin{\cdot}
 \widehat{\boldsymbol{s}}_{ij}
 \right)
 \left(
 \widehat{\boldsymbol{r}}_j\mathbin{\cdot}
 \widehat{\boldsymbol{s}}_{ij}
 \right).
 \label{eq:radial_covariance}
\end{align}
The Cartesian covariance tensor entering this expression is
\begin{equation}
 \Psi_{\alpha\beta}(\boldsymbol{s})
 =
 \Psi_\perp(s)\delta_{\alpha\beta}
 +
 \left[\Psi_\parallel(s)-\Psi_\perp(s)\right]
 \widehat s_\alpha\widehat s_\beta\; ,
 \label{eq:psi_tensor}
\end{equation}
where $\alpha,\beta\in\{x,y,z\}$ denote Cartesian components,
$s=|\boldsymbol{s}|$, $\widehat{\boldsymbol{s}}=\boldsymbol{s}/s$, and
$\delta_{\alpha\beta}$ is the Kronecker delta. The function
$\Psi_\parallel(s)$ is the velocity covariance along the separation
direction, while $\Psi_\perp(s)$ is the covariance in either perpendicular
direction.

We compute $\Psi_\parallel$ and $\Psi_\perp$ from the $z=0$ linear matter
power spectrum in the adopted Planck 2018 cosmology. Their exact integral
definitions and the numerical implementation are given in
Appendix~\ref{app:velocity_covariance}.

Using the fitting operator $A$ in Equation~\eqref{eq:operator}, the
radial velocity covariance $S$ of Equation~\eqref{eq:radial_covariance}
gives the covariance of the fitted parameters,
\begin{equation}
 C_{\rm cv}
 =
 A S A^{\mathsf T},
 \qquad
 C_{\rm pred}
 =
 C_{\rm meas}+C_{\rm cv}.
 \label{eq:cpred}
\end{equation}
The matrix $C_{\rm cv}$ is the cosmic-variance covariance. It describes
how much the fitted monopole and dipole would vary among different
\lcdm\ realizations observed through the same catalog. The matrix
$C_{\rm pred}$ is the total covariance used for the \lcdm\ comparison. It
is the sum of cosmic variance and measurement error.


For two nonoverlapping radial bins $b$ and $b'$, let $S^{bb'}$ be the
matrix of linear radial-velocity covariances between their objects. The
cross-bin cosmic covariance of the fitted moments is
\begin{equation}
 C_{\rm cv}^{bb'}
 =A_b S^{bb'} A_{b'}^{\mathsf T}.
 \label{eq:cross_bin_covariance}
\end{equation}
The measurement cross-covariance is zero because the bins contain
different objects and their measurement errors are treated as independent.
The global tests stack the fitted vectors from all included bins and use
the block matrix formed from Equation~\eqref{eq:cross_bin_covariance} and
the within-bin covariances.

The same construction gives the cosmic cross-covariance of two catalog
windows $a$ and $b$,
\begin{equation}
 C_{\rm cv}^{ab}=A_a S^{ab} A_b^{\mathsf T}.
 \label{eq:cross_catalog_covariance}
\end{equation}
For the direct comparison of two fitted vectors, the covariance of
$\Delta\widehat{\boldsymbol\beta}
=\widehat{\boldsymbol\beta}_a-\widehat{\boldsymbol\beta}_b$ is
\begin{equation}
 C_\Delta
 =C_{\rm pred}^{aa}+C_{\rm pred}^{bb}
 -C_{\rm cv}^{ab}-C_{\rm cv}^{ba},
 \label{eq:difference_covariance}
\end{equation}
where measurement cross-covariance between catalogs is neglected. This
last assumption is used only for the direct catalog-to-catalog comparisons
and is discussed in Section~\ref{sec:limitations}.

\subsection{\texorpdfstring{\lcdm\ $p$-values}{LambdaCDM p-values}}
\label{sec:predictive_statistics}

Select any of the fitted quantities: $\widehat m$,
$\widehat V_{\rm SGX}$, $\widehat V_{\rm SGY}$, and
$\widehat V_{\rm SGZ}$. This may be the monopole alone, the three dipole
components, or all four quantities. Collect the selected values in
$\widehat{\boldsymbol\theta}$, and let $C_\theta$ be the matching
submatrix of $C_{\rm pred}$. We calculate
\begin{equation}
 \begin{aligned}
 Q_\theta
 &=
 \widehat{\boldsymbol\theta}^{\mathsf T}
 C_\theta^{-1}
 \widehat{\boldsymbol\theta},\\
 p_{\lcdm}
 &=
 \Pr\!\left(\chi^2_k\geq Q_\theta\right),
 \end{aligned}
 \label{eq:predictive_p}
\end{equation}
where $k$ is the number of entries in $\widehat{\boldsymbol\theta}$ and
$\chi^2_k$ is a chi-squared random variable with $k$ degrees of freedom.
The value $Q_\theta$ is the squared distance of the best fit from
zero after accounting for the expected variances and correlations. The
$p$-value is the probability, in the adopted \lcdm\ model, of obtaining
$Q_\theta$ at least as large as the observed value. This predictive
statistic is evaluated only after the fit and is not used to determine
$\widehat{\boldsymbol\beta}$.

Unless stated otherwise, the reported $p$-values refer to the joint fit
of $(m,V_{\rm SGX},V_{\rm SGY},V_{\rm SGZ})$. In this parameter order,
the predictive covariance
is partitioned as
\begin{equation}
 C_{\rm pred}
 =
 \begin{pmatrix}
  C_{mm} & C_{mV}\\
  C_{Vm} & C_{VV}
 \end{pmatrix},
 \label{eq:predictive_covariance_blocks}
\end{equation}
where $C_{mm}$ is the monopole variance, $C_{VV}$ is the $3\times3$
dipole covariance, and $C_{mV}=C_{Vm}^{\mathsf T}$ contains their
cross-covariances. Because the joint predictive distribution is Gaussian,
marginalizing means
\begin{align}
 \int d^3\widehat{\boldsymbol V}\,
 \mathcal P(\widehat m,\widehat{\boldsymbol V})
 &= \mathcal N(\widehat m;0,C_{mm}),
 \nonumber\\
 \int d\widehat m\,
 \mathcal P(\widehat m,\widehat{\boldsymbol V})
 &= \mathcal N(\widehat{\boldsymbol V};
 \boldsymbol 0,C_{VV}).
 \label{eq:marginal_predictive_distributions}
\end{align}
Thus, for one radial bin, $p_m$ tests
$\widehat m^2/C_{mm}$ against a chi-squared distribution with one degree
of freedom. The value $p_V$ tests
$\widehat{\boldsymbol V}^{\mathsf T}C_{VV}^{-1}
\widehat{\boldsymbol V}$ against a chi-squared distribution with three
degrees of freedom. The monopole--dipole cross-covariance affects the
joint $p_{m+V}$ test, but it does not enter either marginal PDF. The
omitted fitted quantities are integrated over, not fixed to zero or to
their observed best-fit values. The value
$p_{m+V}$ tests all four fitted quantities jointly and has four degrees
of freedom.

The cross-bin tests use the same definitions after stacking the fitted
values from all included bins. Their $p_m$, $p_V$, and $p_{m+V}$ tests
therefore have $N_{\rm bin}$, $3N_{\rm bin}$, and $4N_{\rm bin}$ degrees
of freedom, respectively.
We emphasize that when a table or figure explicitly states $m=0$,
$p_V$ refers instead to
the separate dipole-only fit in which the monopole is fixed to zero. It
is not the marginalized $p_V$ from the free-$m$ joint fit. None of the
tabulated $p_m$ or $p_V$ values fixes an omitted parameter at its observed
best-fit value. A small $p$-value means that the tested fitted quantities
are uncommon in the
adopted model. It is not the probability that \lcdm\ is false. 
For convenience, we convert this tail probability to the equivalent
two-sided standard-normal significance
$N_\sigma=\Phi^{-1}(1-p/2)$. This conversion provides familiar
astronomical notation. It does not change the degrees of freedom or turn
a multidimensional test into a one-dimensional measurement.

\section{Results}
\label{sec:results}

We organize the \lcdm\ consistency tests at four levels. First,
Table~\ref{tab:joint_results} gives local tests for each catalog and radial
bin. Second, Table~\ref{tab:global_results} combines the assigned bins of
each catalog using their full cross-bin covariance. Third,
Table~\ref{tab:component_results} isolates the monopole and individual
dipole components, while Figure~\ref{fig:component_pulls} shows their
per-bin marginal deviations. Finally,
Sections~\ref{sec:sdss} and~\ref{sec:catalog_influence} examine estimator
choices, calibration, and
local catalog-removal tests. These levels answer different questions and
their probabilities should not be combined.

We use the All Individual four-bin joint test over
$40$--$160\,\hmpc$ as the primary matched-range summary because this is
the common range of the main component catalogs. The six-bin All
Individual test through $300\,\hmpc$ provides the full-range catalog
summary. The per-bin, component-catalog, component, and leave-one-out results
identify the origin and robustness of these summaries and are exploratory
diagnostics. Their probabilities are not adjusted for the number of tests
examined.

\subsection{Fitted monopole and dipole}

\begin{figure*}[t]
\centering
\includegraphics[width=0.92\textwidth]
{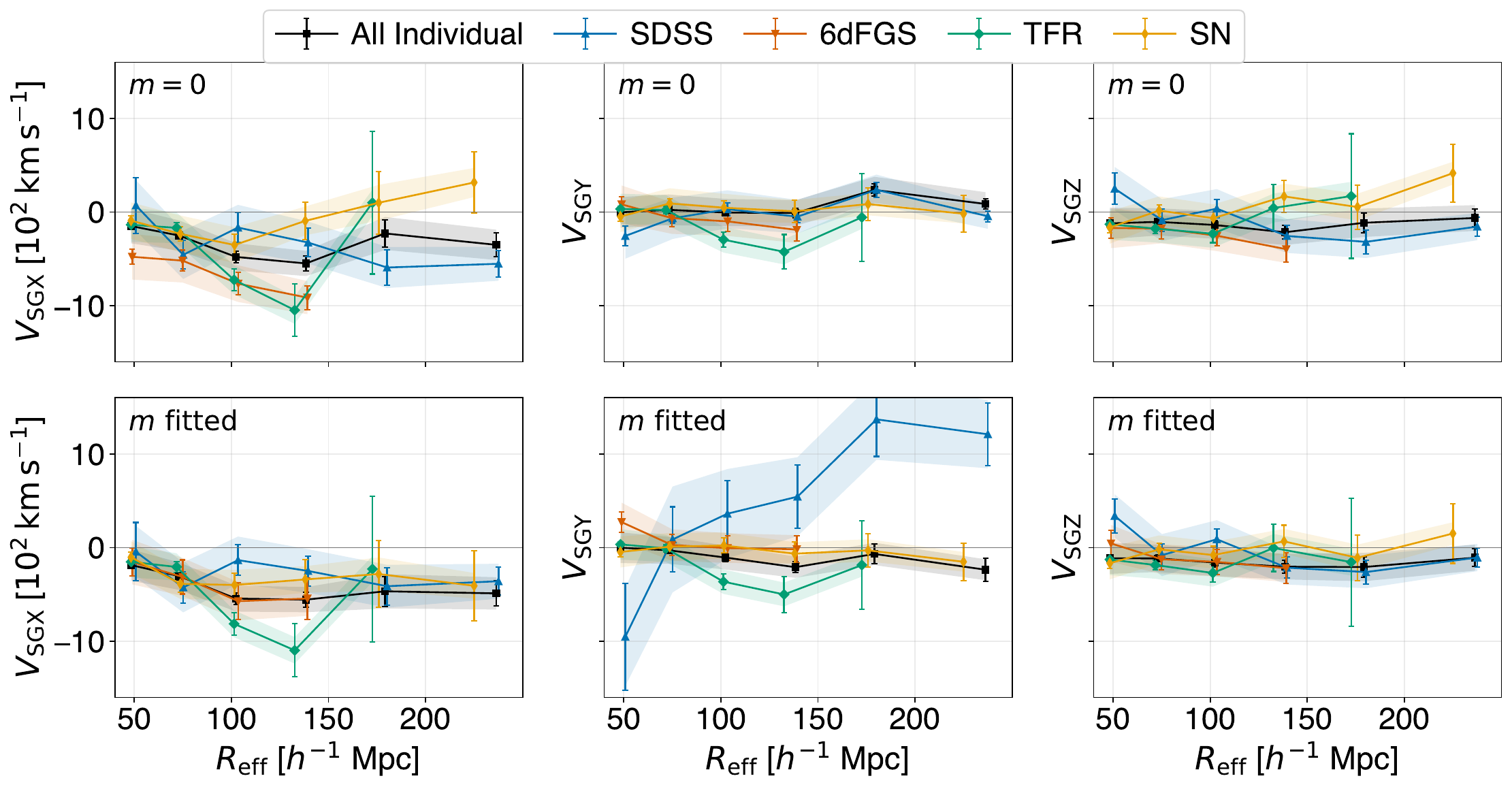}
\caption{Supergalactic dipole components from the hard-bin log-distance
fits. The upper panels fix $m=0$; the lower panels use the joint
$m+\Vdip$ fit. Thin capped error bars show measurement uncertainty; shaded
regions show the separate $1\sigma$ \lcdm\ cosmic variance. Both can differ
between rows because fitting $m$ changes the parameter covariance and the
estimator response. The bands are centered on the measurements only for
comparison, contain no measurement error, and have \lcdm\ mean zero. Units
are $10^2\,\kms$.}
\label{fig:components}
\end{figure*}

Figure~\ref{fig:components} shows the three supergalactic Cartesian
components of the dipole. The top row fixes $m=0$, while the bottom row
fits $m$ jointly with the dipole. Although the catalogs use the same hard
radial bins, each point is placed at the $R_{\rm eff}$ calculated from
that catalog's own objects and inverse-variance weights. The points from
different catalogs in a given bin therefore need not have the same
horizontal coordinate.

The strongest difference between the fixed- and free-monopole fits occurs
in the SDSS SGY component, particularly in the outer bins, while the SGX and SGZ
components are much less affected. Figures~\ref{fig:catalog_angular_overlap}
and~\ref{fig:orientation} show why SGY is the special direction for SDSS:
the sample lies predominantly at positive SGY and has a large positive
weighted mean sightline in that direction. A fitted monopole can therefore
trade strongly against $V_{\rm SGY}$, whereas fixing $m=0$ suppresses this
degeneracy.

The clearest feature in SGX is the TFR catalog. At
$R_{\rm eff}\simeq100\,\hmpc$, both fits give
$V_{\rm SGX}\simeq-8\times10^2\,\kms$, and the component becomes still
more negative in the next bin. At $R_{\rm eff}\simeq130\,\hmpc$, the joint
TFR vector is approximately $(-1097,-501,-4)\,\kms$ in
(SGX, SGY, SGZ). The 6dFGS joint vector in that bin is
$(-544,-18,-220)\,\kms$, while the SDSS vector is
$(-250,544,-215)\,\kms$. Thus TFR and 6dFGS favor negative SGX, whereas
the distinctive SDSS contribution is positive SGY. The All Individual
component curves combine these differently weighted directions.

The upper panels in Figure~\ref{fig:components} come from a
three-parameter dipole fit in which $m$ is fixed exactly to zero. For each
upper-panel component $V_i$, the error bar is the square root of the
corresponding diagonal element of the resulting three-dimensional dipole
covariance; the other two dipole components are marginalized over, but
$m$ is not varied. The upper error bars are therefore the appropriate
measurement uncertainties for the displayed $m=0$ estimates.

The lower panels instead use the full four-parameter joint fit in the
order
$(m,V_{\rm SGX},V_{\rm SGY},V_{\rm SGZ})$. The error bar on each $V_i$
is the square root of its diagonal element in the full joint measurement
covariance. It is therefore marginalized over the fitted monopole and the
other two dipole components: $m$ is allowed to vary with its uncertainty
and covariance with $V_i$. These are not conditional errors calculated
with $m$ held fixed. Incomplete angular coverage produces covariance
between $m$ and the dipole components, so the lower-panel measurement
errors are generally larger than those from the $m=0$ fit.

For each component, the shaded half-width is
$\sqrt{(C_{{\rm cv},VV})_{ii}}$, with the cosmic velocity field propagated
through the same estimator used for the plotted point. Thus the upper
bands use the dipole-only estimator with $m=0$, consistently with the
upper point estimates and error bars, whereas the lower bands use the
joint free-$m$ estimator. The two estimators respond differently to the
same cosmic velocity field, so their band widths can differ. The bands
are centered on the measured curves only as a visual device and contain
no measurement contribution. The total predictive covariance,
$C_{\rm pred}=C_{\rm meas}+C_{\rm cv}$, is used in the \lcdm\ consistency
tests but is not shown in this figure.

\begin{figure*}[t]
\centering
\includegraphics[width=0.90\textwidth]
{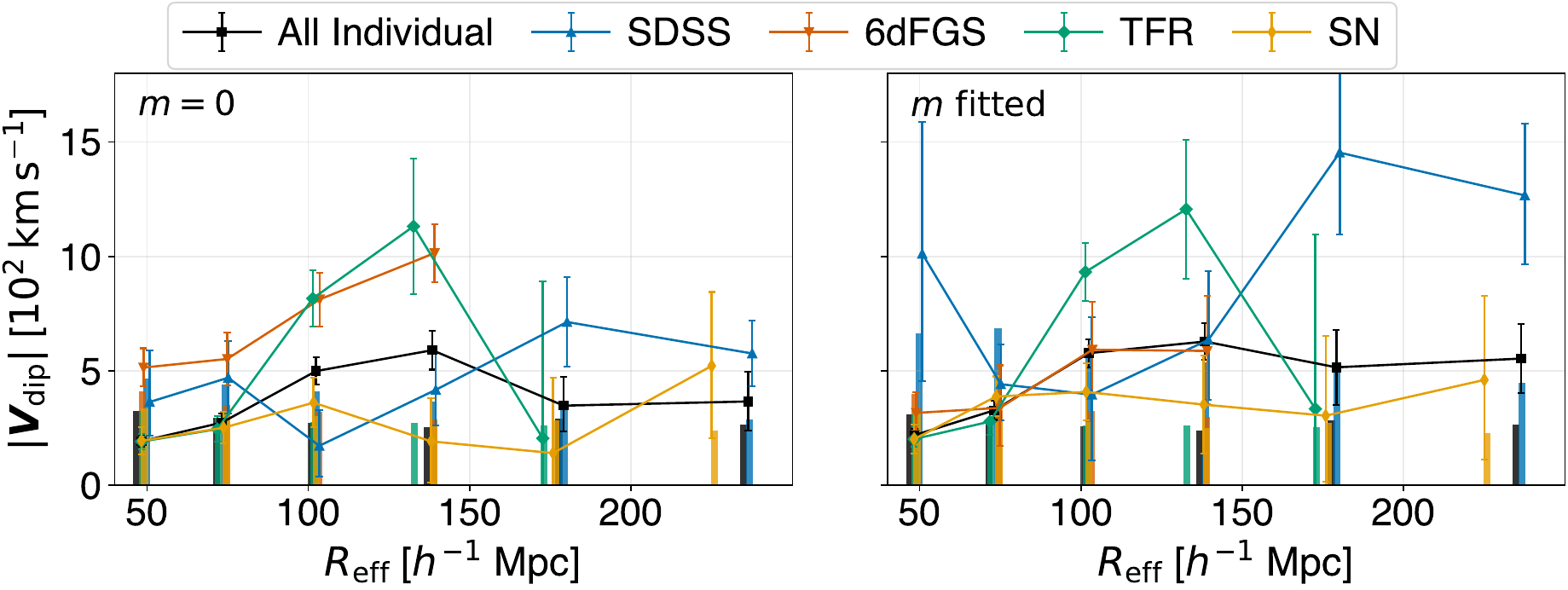}
\caption{Dipole amplitude from the same fits. The left panel fixes
$m=0$. The right panel fits $m$ jointly with the dipole. The plotted unit
is $10^2\,\kms$. Error bars are the generally asymmetric measurement
uncertainties. The shaded
\lcdm\ bars extend from zero to $V_{68}$, the
68.27th percentile of the dipole-amplitude distribution generated by
cosmic variance for each catalog window. They are shown as separate narrow
bars for clarity and are not centered on the measured curves because they
represent the absolute \lcdm\ prediction for $|\Vdip|$, not scatter around
the measurements.}
\label{fig:amplitudes}
\end{figure*}

Figure~\ref{fig:amplitudes} shows curves of
$|\Vdip|=(V_x^2+V_y^2+V_z^2)^{1/2}$. The displayed measurement error is
obtained by propagating the three-component covariance in $C_{\rm meas}$
to the dipole amplitude. Because the amplitude is a nonlinear,
nonnegative quantity, its lower and upper observational errors are
generally asymmetric. For compact numerical summaries in the text, a
single ``$\pm$'' value denotes the corresponding symmetrized
one-standard-deviation uncertainty.

The narrow shaded bars show the dipole amplitudes expected from \lcdm\
cosmic variance. For each catalog and bin, the upper limit $V_{68}$ is
chosen so that 68.27 percent of zero-mean Gaussian dipoles
$\boldsymbol{V}_{\rm cv}$ with covariance $C_{{\rm cv},VV}$ have amplitudes
below it:
\begin{equation}
 \Pr\!\left(
 |\boldsymbol{V}_{\rm cv}|\leq V_{68}
 \right)=0.6827.
 \label{eq:cosmic_amplitude_68}
\end{equation}
The value of $V_{68}$ is calculated from the eigenvalues of
$C_{{\rm cv},VV}$. Each shaded bar extends from zero to $V_{68}$ and
contains 68.27 percent of the predicted amplitudes. We draw separate
narrow bars at each radius for clarity. Unlike the component bands in
Figure~\ref{fig:components}, these bars are not centered on or attached to
the measured curves: they predict the absolute nonnegative quantity
$|\Vdip|$ under \lcdm, rather than scatter around the measured amplitude.
They are model predictions, not uncertainties on the measurements.

The dipole amplitudes in the left and right panels of
Figure~\ref{fig:amplitudes} come from different fits. The left panel fixes
$m=0$, whereas the right panel fits $m$ jointly with the dipole. Fixing
$m=0$ removes the covariance between the monopole and dipole, while
fitting $m$ allows angular coverage to couple them. This distinction is
large for SDSS because its window strongly
couples $m$ to $V_{\rm SGY}$. In the $200$--$300\,\hmpc$ bin, the fixed
fit gives $|\Vdip|=576\pm144\,\kms$. Fitting $m$ to the same data gives
$m=-1046\pm274\,\kms$ and
$|\Vdip|=1267\pm308\,\kms$, driven mainly by
$V_{\rm SGY}=1210\,\kms$.

SDSS provides 18,934 of the 19,101 All Individual objects in this bin,
but the two curves need not coincide because the fits depend on the
statistical weights and angular coverage, not simply on object counts.
An object-matched test gives nearly identical amplitudes whether the SDSS
objects use their native or All Individual ${\rm DM}/e{\rm DM}$ values. The choice of
distance-modulus source is therefore not responsible for the difference.

The 167 non-SDSS objects have disproportionate influence. In particular,
the 115 SN have a median $e{\rm DM}\simeq0.15$ mag, compared with about
$0.52$ mag for the full outer-bin sample. Removing SN brings the
fixed-$m$ amplitude close to the SDSS value, but the free-$m$ amplitude
remains lower because the remaining 42 TFR and 10 Other FP objects change
the angular window. The SDSS free fit has a strong monopole--SGY
correlation, $\rho_{\rm meas}(m,V_{\rm SGY})\simeq-0.98$, so this window
change shifts $m$ and $V_{\rm SGY}$ while affecting SGX and SGZ much less.
This explains why the fixed-$m$ curves agree but the free-$m$ curves
remain separated in Figures~\ref{fig:amplitudes}
and~\ref{fig:components}.

In the $120$--$160\,\hmpc$ bin, the All Individual joint fit gives
\begin{equation}
 \widehat m=275\pm43\,\kms,
 \qquad
 |\widehat{\Vdip}|=628\pm82\,\kms
 \label{eq:all_individual_outer_result}
\end{equation}
at $R_{\rm eff}\simeq140\,\hmpc$. The quoted uncertainties are measurement
uncertainties obtained from $C_{\rm meas}$.

Among the component catalogs, TFR has the largest best-fit joint dipole,
$1206\pm303\,\kms$ at $R_{\rm eff}\simeq130\,\hmpc$. Its large amplitude
does not by itself determine its influence on All Individual because the
component catalogs have different sizes, weights, and angular windows. A
similar scalar amplitude also need not imply a similar direction.

\subsection{\texorpdfstring{\lcdm}{LambdaCDM} consistency levels}

The dipole amplitude alone does not determine consistency with \lcdm,
because the predicted covariance depends on the catalog window and the
fitted monopole can be correlated with the dipole. We therefore quantify
the comparison with the $p_m$, $p_V$, and $p_{m+V}$ tests defined in
Section~\ref{sec:predictive_statistics}. Table~\ref{tab:joint_results} gives the
per-bin results. The number in parentheses beside each $p$-value is its
equivalent two-sided Gaussian deviation $N_\sigma$; it is a familiar way
to quote the same tail probability and does not change the dimensionality
of the test. Table~\ref{tab:global_results} then combines the radial bins
of each catalog using their full cross-bin covariance.

\begin{table*}[t]
\centering
\scriptsize
\setlength{\tabcolsep}{2.5pt}
\caption{\lcdm\ $p$-values in each bin}
\label{tab:joint_results}
\begin{tabular}{lrrrrrrr}
\toprule
Catalog &
Bin &
$R_{\rm eff}$ &
$N_{\rm obj}$ &
$p_m$ ($N_\sigma$) &
$p_V$ ($N_\sigma$; free $m$) &
$p_{m+V}$ ($N_\sigma$; free $m$) &
$\chi^2_{\rm GOF}$ \\
&
($\hmpc$) &
($\hmpc$) &
&
&
&
&
\\
\midrule
All Individual & 40--60   & 48  & 4,239 & 0.630 (0.48) & 0.646 (0.46) & 0.753 (0.31) & 4418.9 \\
SDSS           & 40--60   & 51  & 210   & 0.356 (0.92) & 0.515 (0.65) & 0.587 (0.54) & 225.1 \\
6dFGS          & 40--60   & 49  & 559   & 0.203 (1.27) & 0.448 (0.76) & 0.287 (1.06) & 488.7 \\
TFR            & 40--60   & 48  & 3,180 & 0.817 (0.23) & 0.706 (0.38) & 0.833 (0.21) & 3274.3 \\
SN             & 40--60   & 48  & 176   & 0.905 (0.12) & 0.752 (0.32) & 0.877 (0.15) & 224.3 \\
Other FP       & 40--60   & 48  & 230   & 0.317 (1.00) & $\mathbf{0.00166\ (3.15)}$ & $\mathbf{0.00155\ (3.17)}$ & 341.2 \\
\addlinespace
All Individual & 60--90   & 73  & 6,585 & 0.466 (0.73) & 0.197 (1.29) & 0.275 (1.09) & 5912.2 \\
SDSS           & 60--90   & 75  & 1,567 & 0.797 (0.26) & 0.552 (0.59) & 0.635 (0.47) & 1605.6 \\
6dFGS          & 60--90   & 75  & 1,156 & 0.445 (0.76) & 0.584 (0.55) & 0.338 (0.96) & 986.6 \\
TFR            & 60--90   & 72  & 3,451 & 0.613 (0.51) & 0.382 (0.87) & 0.504 (0.67) & 3158.9 \\
SN             & 60--90   & 74  & 238   & 0.287 (1.06) & 0.202 (1.28) & 0.277 (1.09) & 307.7 \\
Other FP       & 60--90   & 75  & 371   & 0.589 (0.54) & 0.311 (1.01) & 0.370 (0.90) & 302.6 \\
\addlinespace
All Individual & 90--120  & 102 & 6,767 & 0.148 (1.45) & $\mathbf{0.00208\ (3.08)}$ & $\mathbf{0.00209\ (3.08)}$ & 5798.6 \\
SDSS           & 90--120  & 103 & 2,583 & 0.560 (0.58) & 0.828 (0.22) & 0.924 (0.10) & 2624.5 \\
6dFGS          & 90--120  & 104 & 2,097 & 0.428 (0.79) & 0.179 (1.34) & 0.0179 (2.37) & 1506.6 \\
TFR            & 90--120  & 101 & 1,654 & 0.0925 (1.68) & $\mathbf{2.54\times10^{-5}\ (4.21)}$ & $\mathbf{1.94\times10^{-5}\ (4.27)}$ & 1682.3 \\
SN             & 90--120  & 102 & 178   & 0.724 (0.35) & 0.219 (1.23) & 0.352 (0.93) & 242.7 \\
Other FP       & 90--120  & 103 & 440   & 0.276 (1.09) & 0.125 (1.53) & 0.168 (1.38) & 230.6 \\
\addlinespace
All Individual & 120--160 & 138 & 8,481 & 0.0148 (2.44) & $\mathbf{6.67\times10^{-4}\ (3.40)}$ & $\mathbf{1.11\times10^{-4}\ (3.87)}$ & 6729.4 \\
SDSS           & 120--160 & 140 & 4,606 & 0.258 (1.13) & 0.401 (0.84) & 0.463 (0.73) & 4295.3 \\
6dFGS          & 120--160 & 139 & 3,097 & 0.135 (1.49) & 0.207 (1.26) & $\mathbf{4.36\times10^{-4}\ (3.52)}$ & 2187.1 \\
TFR            & 120--160 & 133 & 501   & 0.282 (1.08) & 0.00435 (2.85) & 0.00849 (2.63) & 526.4 \\
SN             & 120--160 & 138 & 108   & 0.0141 (2.45) & 0.602 (0.52) & 0.137 (1.49) & 152.2 \\
Other FP       & 120--160 & 134 & 254   & 0.122 (1.55) & 0.241 (1.17) & 0.138 (1.48) & 165.2 \\
\addlinespace
SDSS           & 160--200 & 180 & 6,099 & 0.0415 (2.04) & 0.0171 (2.38) & 0.0375 (2.08) & 5639.9 \\
SN             & 160--200 & 176 & 93    & 0.00881 (2.62) & 0.833 (0.21) & 0.138 (1.48) & 95.7 \\
SDSS           & 200--300 & 238 & 18,934 & 0.00740 (2.68) & 0.0134 (2.47) & 0.0104 (2.56) & 15965.0 \\
\bottomrule
\end{tabular}
\tablecomments{Results use the four-parameter $m+\Vdip$ fit. Parentheses
give the equivalent two-sided $N_\sigma$; unadjusted values of at least
$3\sigma$ are bold. The final column gives the measurement-only residual
goodness-of-fit statistic $\chi^2_{\rm GOF}$; its degrees of freedom are
$N_{\rm obj}-4$. It is distinct from the \lcdm\ predictive tests. Catalog
rows overlap and are not independent.}
\end{table*}

\begin{table*}[t]
\centering
\scriptsize
\setlength{\tabcolsep}{5.0pt}
\caption{Cross-bin \lcdm\ $p$-values over catalog-specific bins}
\label{tab:global_results}
\begin{tabular}{lrrrrrr}
\toprule
Catalog &
Bin &
$N_{\rm bin}$ &
$N_{\rm obj}$ &
$p_m$ ($N_\sigma$) &
$p_V$ ($N_\sigma$; free $m$) &
$p_{m+V}$ ($N_\sigma$; free $m$) \\
&
($\hmpc$) &
&
&
&
&
\\
\midrule
All Individual & 40--160 & 4 & 26,072 & 0.185 (1.33) & 0.0195 (2.34) & 0.0156 (2.42) \\
All Individual & 40--300 & 6 & 51,520 & 0.143 (1.47) & 0.0222 (2.29) & 0.0181 (2.36) \\
SDSS           & 40--300 & 6 & 33,999 & 0.0617 (1.87) & 0.0954 (1.67) & 0.0924 (1.68) \\
6dFGS          & 40--160 & 4 & 6,909  & 0.464 (0.73) & 0.571 (0.57) & 0.0512 (1.95) \\
TFR            & 40--160 & 4 & 8,786  & 0.444 (0.77) & $\mathbf{5.74\times10^{-4}\ (3.44)}$ & $\mathbf{0.00135\ (3.21)}$ \\
Other FP       & 40--160 & 4 & 1,295  & 0.203 (1.27) & 0.0110 (2.54) & 0.00529 (2.79) \\
SN             & 40--200 & 5 & 793    & 0.0253 (2.24) & 0.820 (0.23) & 0.315 (1.00) \\
\bottomrule
\end{tabular}
\tablecomments{Each row combines the listed bins using their full
cross-bin covariance. Parentheses give the equivalent two-sided
$N_\sigma$; unadjusted values of at least $3\sigma$ are bold. The ranges
differ, and cross-catalog covariance is not included.}
\end{table*}

The clearest local deviations occur in the $90$--$120$ and
$120$--$160\,\hmpc$ bins. In the former, the TFR dipole deviation is
$4.21\sigma$ and the All Individual dipole deviation is $3.08\sigma$.
In the latter, the All Individual dipole and joint deviations are
$3.40\sigma$ and $3.87\sigma$. Other FP also has an isolated
dipole and joint deviation of $3.15\sigma$ and $3.17\sigma$ in the
$40$--$60\,\hmpc$ bin, but this feature does not persist across its four
bins.

These local deviations do not produce a comparably strong result when the
four common bins spanning $40$--$160\,\hmpc$ are tested together. For All
Individual, the cross-bin
values are $p_m=0.185$ ($1.33\sigma$), $p_V=0.0195$ ($2.34\sigma$), and
$p_{m+V}=0.0156$ ($2.42\sigma$). The combined catalog therefore shows a
modest full radial-range deviation, with no cross-bin result above
$3\sigma$.
Extending All Individual through $300\,\hmpc$ gives
$p_m=0.143$ ($1.47\sigma$), $p_V=0.0222$ ($2.29\sigma$), and
$p_{m+V}=0.0181$ ($2.36\sigma$). Thus, adding the two outer bins does not
strengthen the combined-bin deviation. Among the component-catalog rows
covering the common $40$--$160\,\hmpc$ range, TFR gives
dipole and joint deviations of $3.44\sigma$ and $3.21\sigma$, while
Other FP gives $2.54\sigma$ and $2.79\sigma$. Rows covering different
ranges should not be ranked directly.

Figure~\ref{fig:lcdm_methods} provides the corresponding 68.3 percent
measurement and predictive ellipses for the two main bins. These contours
are useful visualizations of the catalog-dependent covariances, but the
probabilities in Tables~\ref{tab:joint_results}
and~\ref{tab:global_results} use all fitted dimensions. We next identify the
components responsible for the smallest probabilities.

\subsection{Which components drive the small \texorpdfstring{$p$}{p}-values?}

The single $p_V$ value combines SGX, SGY, and SGZ. It does not show which
direction causes a small value. We therefore test each component across
all assigned bins. These tests include the correlations between bins.
Table~\ref{tab:component_results} shows that SGX is the least consistent
direction for All Individual. Its SGY and SGZ results are consistent with
\lcdm, with SGZ the most consistent component. The monopole result is also
consistent.

\begin{figure*}[t]
\centering
\includegraphics[width=0.94\textwidth]
{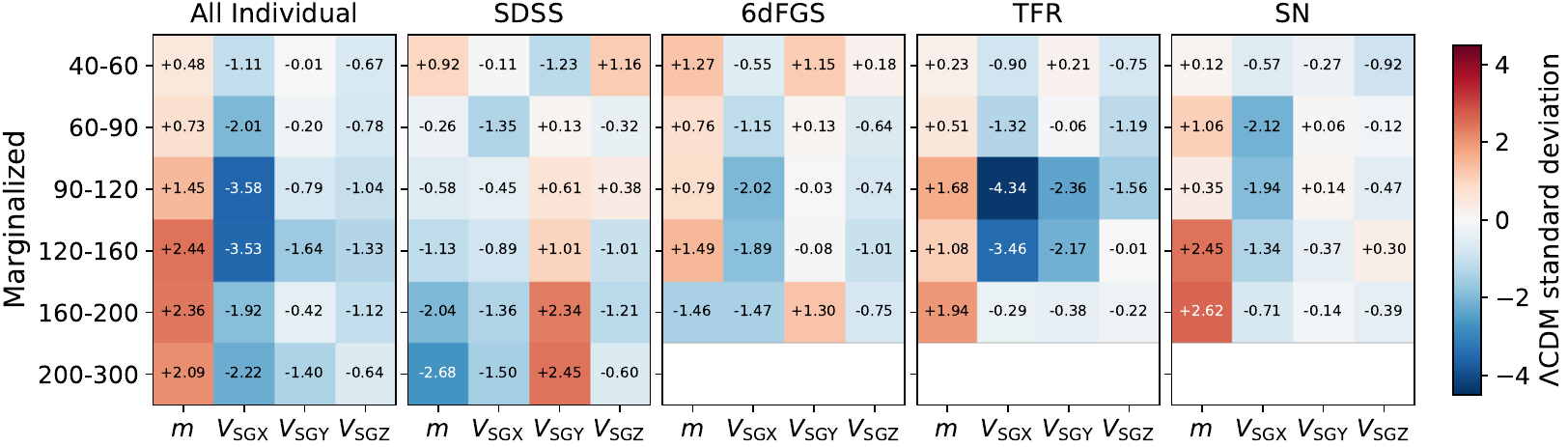}
\caption{Standardized component deviations for All Individual and the
component catalogs. Columns show $m$, SGX, SGY, and SGZ; rows show bins.
Each cell is the signed marginal deviation from the zero \lcdm\ mean in
predictive-standard-deviation units. Red is positive and blue is negative;
blank cells are unused. The cells are correlated.}
\label{fig:component_pulls}
\end{figure*}

For a given catalog, let $\widehat{\boldsymbol{\beta}}$ collect all fitted
values of $m$, SGX, SGY, and SGZ in its radial bins, and let
$C=C_{\rm pred}$ be their joint \lcdm\ covariance, including measurement
error. For an individual entry $\widehat\beta_i$, its signed marginal
deviation is
\begin{equation}
 z_{i,{\rm marg}}
 =\frac{\widehat\beta_i}{\sqrt{C_{ii}}}.
 \label{eq:marginal_deviation}
\end{equation}
All other fitted quantities are integrated over according to the joint
covariance; they are not fixed to zero. Each cell in
Figure~\ref{fig:component_pulls} is therefore a one-dimensional,
one-degree-of-freedom comparison. The cells are correlated and cannot be
combined as independent tests.

In the All Individual panel, the largest deviations occur in SGX in the
$90$--$120$ and $120$--$160\,\hmpc$ bins and are $-3.58\sigma$ and
$-3.53\sigma$. The two dark-blue cells therefore identify negative SGX
as the main source of the local All Individual deviation. TFR shows the
same component and sign in these bins. The formal cross-bin component
tests over the four primary bins are given in
Table~\ref{tab:component_results}.

\begin{table*}[t]
\centering
\scriptsize
\setlength{\tabcolsep}{7.0pt}
\caption{Cross-bin \lcdm\ tests of individual physical components}
\label{tab:component_results}
\begin{tabular}{lrrrrrr}
\toprule
Catalog & $p_m$ & $p_{\mathrm{SGX}}$ & $p_{\mathrm{SGY}}$ &
$p_{\mathrm{SGZ}}$ & Least consistent & Most consistent \\
\midrule
All Individual & 0.185 & 0.00108 & 0.604 & 0.661 & SGX & SGZ \\
SDSS           & 0.0617 & 0.382 & 0.0601 & 0.604 & SGY & SGZ \\
6dFGS          & 0.464 & 0.152 & 0.833 & 0.818 & SGX & SGY \\
TFR            & 0.444 & $2.13\times10^{-5}$ & 0.0506 & 0.572 & SGX & SGZ \\
Other FP       & 0.203 & 0.00814 & 0.106 & 0.0758 & SGX & $m$ \\
SN             & 0.0253 & 0.196 & 0.996 & 0.890 & $m$ & SGY \\
\bottomrule
\end{tabular}
\tablecomments{Each $p$-value combines one component across the assigned
bins, including cross-bin covariance and marginalizing over the other
fitted components. The component tests are correlated.}
\end{table*}

\subsection{Measurement and predictive contours}

\begin{figure*}[t]
\centering
\includegraphics[width=0.84\textwidth]
{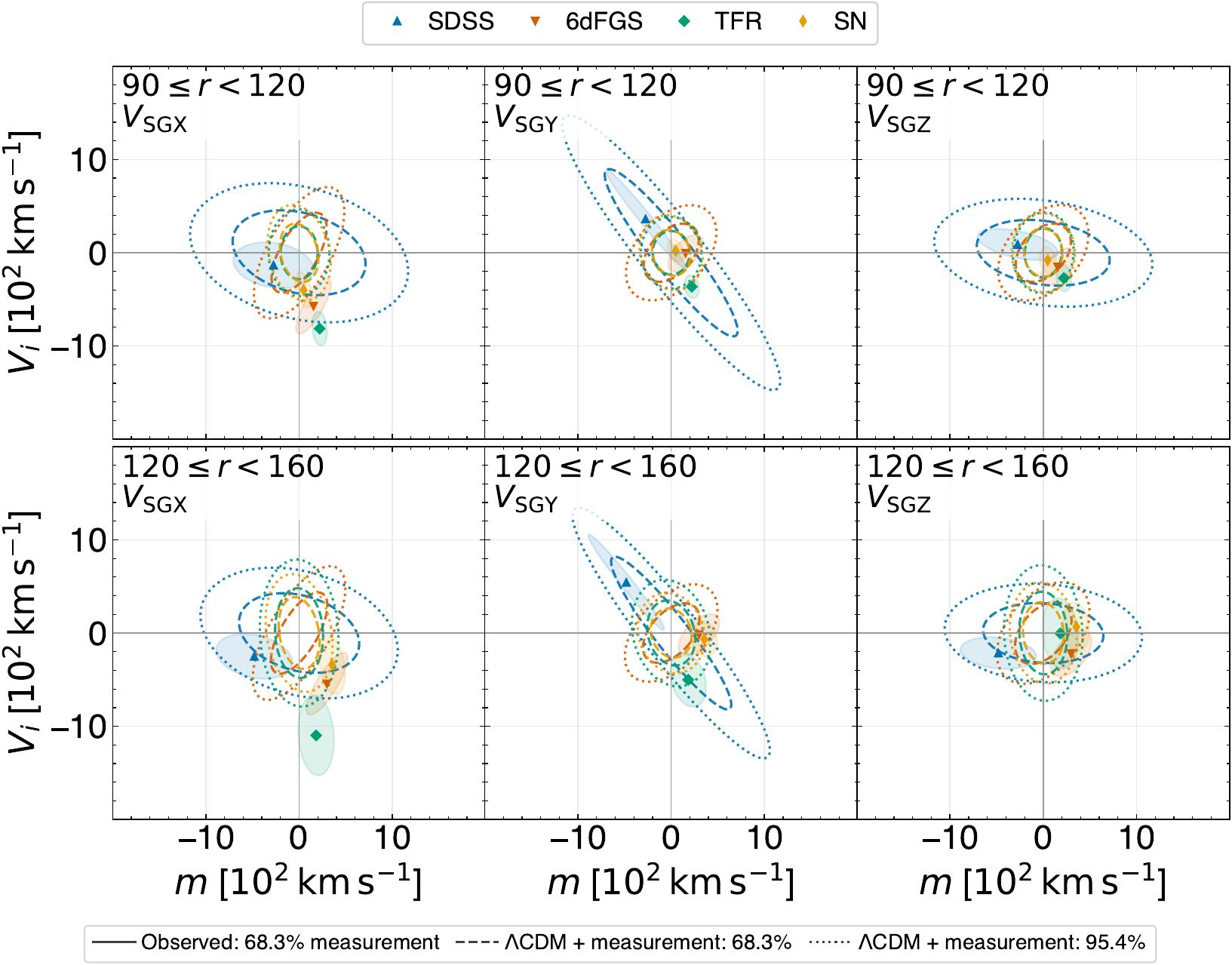}
\caption{Measurement errors and \lcdm\ expectations for the component
catalogs in the $90$--$120$ and $120$--$160\,\hmpc$ bins. Each filled
ellipse is the 68.3 percent measurement region around the best fit.
Dashed and dotted ellipses are the 68.3 and 95.4 percent predictive
regions around zero, including measurement error and cosmic variance for
each catalog window. Reported $p$-values use the full fitted covariance,
not these two-dimensional projections. Axes use $10^2\,\kms$.}
\label{fig:lcdm_methods}
\end{figure*}

Figure~\ref{fig:lcdm_methods} compares the measurement errors with the
range expected in repeated \lcdm\ realizations. The \lcdm\ range is
centered on zero. It includes both cosmic variance and measurement error.
It differs between catalogs because each catalog has different object
positions, sky coverage, and weights. The figure shows the two main
contour bins for SDSS, 6dFGS, TFR, and SN. Other FP remains in the
numerical tables but is omitted from the confidence figures for clarity.
The contours provide a visual counterpart to the probabilities in
Table~\ref{tab:joint_results}, but the formal tests use all relevant
dimensions and are not inferred from the visible overlap of ellipses.

In the $120$--$160\,\hmpc$ bin, the SDSS dipole and joint deviations are
only $0.84\sigma$ and $0.73\sigma$. Its large scalar amplitude is
therefore ordinary under the broad predictive covariance associated with
its window.

The 6dFGS result illustrates why the correlations matter. Its separate
monopole and dipole deviations are $1.49\sigma$ and $1.26\sigma$, while
their joint deviation is $3.52\sigma$. The joint value is larger because
a particular combination of the monopole and dipole is predicted much
more tightly than either one alone. In the 6dFGS \lcdm\ covariance,
the approximate correlations of $m$ with SGX, SGY, and SGZ are 0.68, 0.49,
and 0.41. The observed positive monopole and negative SGX component probe
a combination that is more unusual than either separate test.

The SN sample provides a useful independent window check. In the
$90$--$120\,\hmpc$ bin, its dipole and joint deviations are
$1.23\sigma$ and $0.93\sigma$. In the $120$--$160\,\hmpc$ bin, they are
$0.52\sigma$ and $1.49\sigma$. The fitted SN monopole in the latter bin
is a larger $2.45\sigma$ deviation, but its dipole amplitude is consistent
with the predictive covariance.

\section{The SDSS result through \texorpdfstring{$300\,\hmpc$}{300 Mpc/h}}
\label{sec:sdss}

SDSS is the only component catalog with useful support beyond
$160\,\hmpc$. It contains 6,099 accepted objects in the
$160$--$200\,\hmpc$ bin and 18,934 in the
$200$--$300\,\hmpc$ bin. We first establish the result through
$160\,\hmpc$, where the four common bins can be tested together. We then
examine the two outer SDSS bins.

\subsection{Fixed and free monopole fits}
\label{sec:sdss_joint_fit}

We compare two estimators. The first fixes $m=0$ and fits only
$\Vdip$. The second fits
$\widehat{\boldsymbol\beta}
=(m,V_{\rm SGX},V_{\rm SGY},V_{\rm SGZ})$.
The fixed-$m$ estimator is recomputed consistently. It is not obtained
by dropping $m$ from the free fit. If $N$ is the measurement covariance
of the free estimator, then
\begin{equation}
 \widehat{\boldsymbol V}_{m=0}
 =
 \widehat{\boldsymbol V}
 -N_{Vm}N_{mm}^{-1}\widehat m .
 \label{eq:sdss_fixed_monopole}
\end{equation}
We apply the same linear transformation to the measurement and \lcdm\
cross-bin covariance matrices.

Across the four SDSS bins from $40$ to $160\,\hmpc$, the fixed-$m$
dipole deviation is $0.31\sigma$. The free fit gives $0.36\sigma$ for
the three dipole components and $0.25\sigma$ for all four parameters.
The SDSS data through $160\,\hmpc$ are therefore fully consistent with
the adopted \lcdm\ model under either treatment of the monopole.
The largest difference between the two estimators occurs in SGY in the
two outer bins. The free-$m$ estimates become large and positive, while
the fixed-$m$ estimates remain near zero. This behavior anticipates the
strong outer-bin $m$--SGY degeneracy shown below.

\subsection{The two outer bins}
\label{sec:sdss_outer}

For the free fit, the data-only outer-bin estimates are
\begin{align}
\widehat{\boldsymbol\beta}_{160-200}
 &=(-956,-413,1369,-261)\,\kms ,
 \nonumber\\
\widehat{\boldsymbol\beta}_{200-300}
 &=(-1046,-359,1210,-107)\,\kms .
\label{eq:sdss_outer_fits}
\end{align}
The strong negative monopoles and positive SGY components are highly
correlated because most SDSS sightlines point toward positive SGY.
Table~\ref{tab:sdss_outer_models} shows that the inferred significance
depends on whether the monopole is fixed.

\begin{table}[t]
\centering
\scriptsize
\setlength{\tabcolsep}{2.4pt}
\caption{\lcdm\ consistency of the outer SDSS bins}
\label{tab:sdss_outer_models}
\begin{tabular}{@{}lrrr@{}}
\toprule
Bin &
\multicolumn{1}{c}{$p_V$ ($N_\sigma$)} &
\multicolumn{1}{c}{$p_V$ ($N_\sigma$)} &
\multicolumn{1}{c}{$p_{m+V}$ ($N_\sigma$)} \\
($\hmpc$) &
\multicolumn{1}{c}{$m=0$} &
\multicolumn{1}{c}{free $m$} &
\multicolumn{1}{c}{free $m$} \\
\midrule
160--200 & 0.1055 (1.62) & 0.0171 (2.38) & 0.0375 (2.08) \\
200--300 & 0.0995 (1.65) & 0.0134 (2.47) & 0.0104 (2.56) \\
160--300 & 0.0590 (1.89) & 0.00768 (2.67) & 0.00601 (2.75) \\
\bottomrule
\end{tabular}
\tablecomments{The combined row includes the covariance between bins.
Parentheses give the equivalent two-sided $N_\sigma$.}
\end{table}

With $m=0$, the covariance-aware combination of the two outer-bin
dipoles differs from \lcdm\ by $1.89\sigma$. When $m$ is free, the
combined dipole and full monopole--dipole deviations are $2.67\sigma$
and $2.75\sigma$. The contrast is a consequence of the SDSS angular
window and the strong $m$--SGY degeneracy. It demonstrates that the
outer-bin assessment is model dependent.

\begin{figure}[t]
\centering
\includegraphics[width=\columnwidth]
{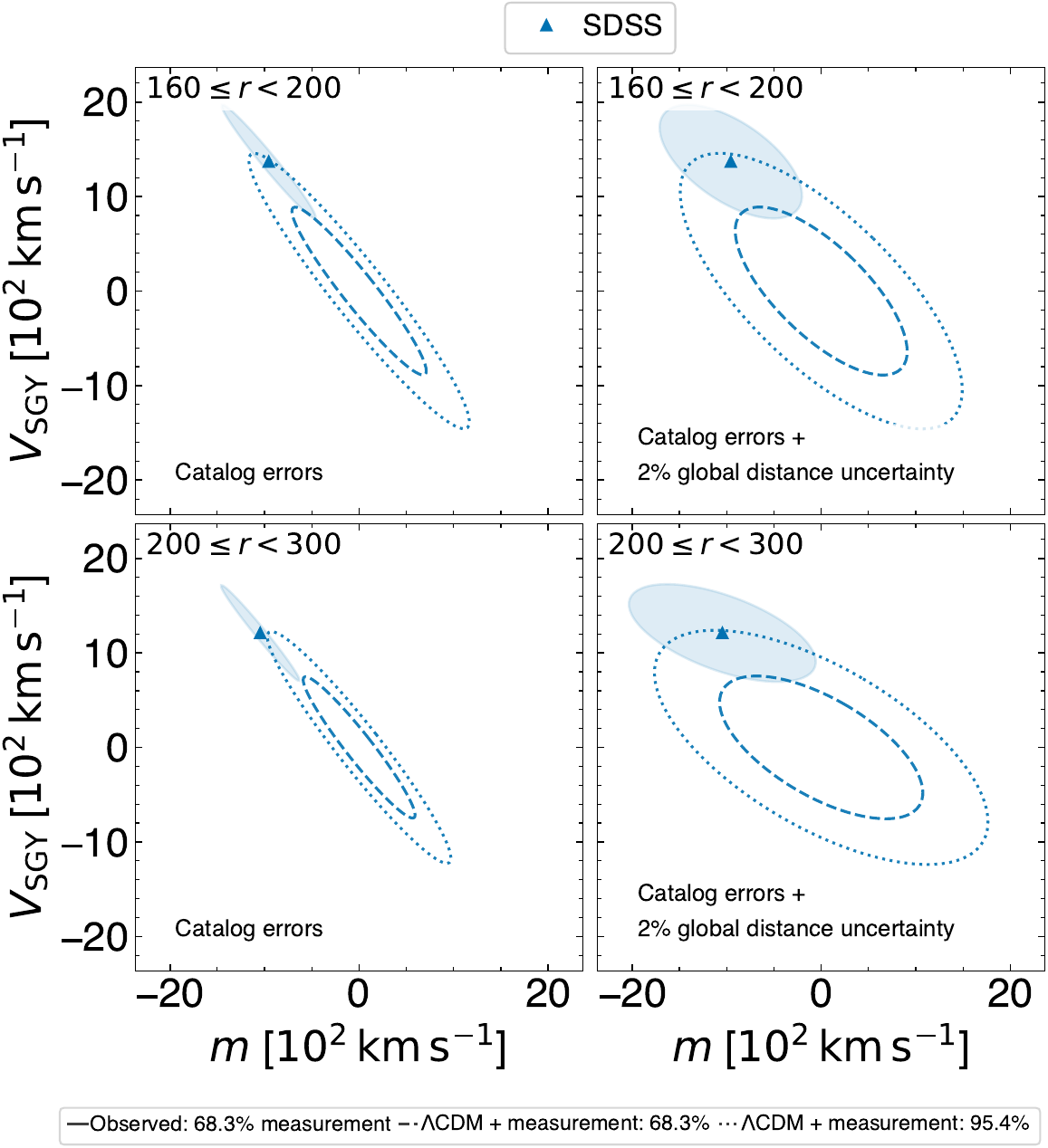}
\caption{Marginal $m$--SGY contours for the two outer SDSS bins. The left
column includes catalog errors only; the right column also includes an
illustrative shared two-percent distance-scale uncertainty. Triangles mark
the best fits, shaded regions show the 68.3 percent measurement
constraints, and dashed and dotted contours show the 68.3 and 95.4
percent \lcdm\ predictive regions, respectively.}
\label{fig:sdss_zero_point_contours}
\end{figure}

Figure~\ref{fig:sdss_zero_point_contours} illustrates the effect of adding
a shared two-percent SDSS distance-scale uncertainty. It broadens the
$m$--SGY constraints and reduces the combined outer-bin joint deviation
from $2.75\sigma$ to $2.39\sigma$, while leaving the dipole deviation
nearly unchanged at $2.66\sigma$. Across all six SDSS bins, the joint
deviation decreases from $1.68\sigma$ to $1.34\sigma$. The assumed
two-percent value is a sensitivity test, not a new calibration
measurement.

\subsection{Quadrupole and angular residuals}
\label{sec:sdss_quadrupole}

We test whether coherent angular structure beyond a dipole changes the
outer-bin interpretation. The extended model is
\begin{equation}
 u_{\ln,i}=m+\widehat{\boldsymbol r}_i\cdot\Vdip
 +\widehat{\boldsymbol r}_i^{\mathsf T}
 Q\widehat{\boldsymbol r}_i ,
 \label{eq:sdss_quadrupole}
\end{equation}
where $Q$ is symmetric and trace free and has five independent
components, each with velocity units. Table~\ref{tab:sdss_quadrupole}
tests whether adding these five components significantly improves the
$m+\Vdip$ fit. The reduction $\Delta\chi^2$ in the minimum chi-squared is
compared with a chi-squared distribution with five degrees of freedom.

\begin{table}[t]
\centering
\scriptsize
\caption{Quadrupole tests in the outer SDSS bins}
\label{tab:sdss_quadrupole}
\begin{tabular}{@{}lrr@{}}
\toprule
Bin &
\multicolumn{1}{c}{$\Delta\chi^2$, add $Q$} &
\multicolumn{1}{c}{$p$, add $Q$} \\
($\hmpc$) & & \\
\midrule
160--200 & 4.43 & 0.489 \\
200--300 & 3.69 & 0.595 \\
160--300 & 4.97 & 0.420 \\
\bottomrule
\end{tabular}
\tablecomments{The likelihood-ratio test adds the five components of
$Q$ to $m+\Vdip$.}
\end{table}

The quadrupole does not improve the fit significantly in any of the
listed ranges. The visible outer-bin pattern therefore does not require
angular structure beyond the monopole and dipole in these data.
The main outer-bin issue remains the coherent, window-dependent
monopole--dipole combination.

\section{Component-catalog influence within All Individual}
\label{sec:catalog_influence}

The component catalogs have little pairwise overlap, but all are subsets of
All Individual. We remove from All Individual every object belonging to
the selected component catalog, refit the joint model, and recompute both
covariance terms for the remaining positions and weights. Because catalog
memberships overlap, these removals are not additive: an object removed
through one catalog can also belong to another component catalog.
Table~\ref{tab:leave_one_out} shows the two
discrepant radial bins. These are local sensitivity tests of the bump,
not cross-bin tests of the global consistency of each reduced catalog.

At $90$--$120\,\hmpc$, removing TFR gives the largest change. It
changes SGX from $-545$ to $-461\,\kms$ and reduces the dipole deviation
from $3.08\sigma$ to $2.20\sigma$. Removing 6dFGS gives
$2.43\sigma$. Removing SDSS makes the SGX estimate more negative and
raises the dipole deviation to $3.60\sigma$.
Removing SN also raises the amplitude, from $578$ to $600\,\kms$, and
gives dipole and joint deviations of $3.16\sigma$ and $3.34\sigma$.
Removing TFR or 6dFGS lowers the All Individual amplitude in this bin,
whereas removing SDSS or SN raises it.

At $120$--$160\,\hmpc$, 6dFGS has the decisive influence. Removing 6dFGS
changes SGX from $-556$ to $-411\,\kms$ and reduces the dipole deviation
from $3.40\sigma$ to $1.56\sigma$. The joint deviation falls from
$3.87\sigma$ to $1.51\sigma$. Removing TFR produces a smaller change and
leaves a large joint deviation. Removing Other FP changes almost nothing.
The large TFR-only amplitude is therefore not evidence that TFR alone
drives the outer All Individual result. The larger 6dFGS sample and fit
weight, together with its angular window, give it greater influence on
the combined fit in this bin.
Removing SN has the opposite effect. It raises the amplitude from $628$
to $724\,\kms$ and raises the dipole and joint deviations to
$3.77\sigma$ and $4.11\sigma$. Removing SDSS similarly raises the
amplitude to $691\,\kms$. Thus, the bump at
$R_{\rm eff}\simeq100$--$140\,\hmpc$ reflects opposing changes under
catalog removal. The upper-bin result is most sensitive to 6dFGS because
only its removal lowers both the dipole and joint deviations below
$2\sigma$. This sensitivity is not an additive decomposition of the
combined signal, since every removal also changes the survey window and
predictive covariance.

We also compare the standalone SN and 6dFGS joint fits directly. For
their difference, the covariance is given by
Equation~\eqref{eq:difference_covariance} and includes both measurement
covariances and the cross-window \lcdm\ cosmic covariance. The dipole-difference $p$-values
are 0.946 and 0.643 in the $90$--$120$ and
$120$--$160\,\hmpc$ bins. The corresponding four-parameter values are
0.721 and 0.796. The two catalogs are therefore mutually consistent in
these bins despite pulling the combined All Individual fit in opposite
directions. The calculation treats measurement errors as independent.
There are only six shared objects in the lower bin and one in the upper bin,
so the omitted measurement cross-covariance is expected to be small.

\begin{figure*}[t]
\centering
\includegraphics[width=0.92\textwidth]
{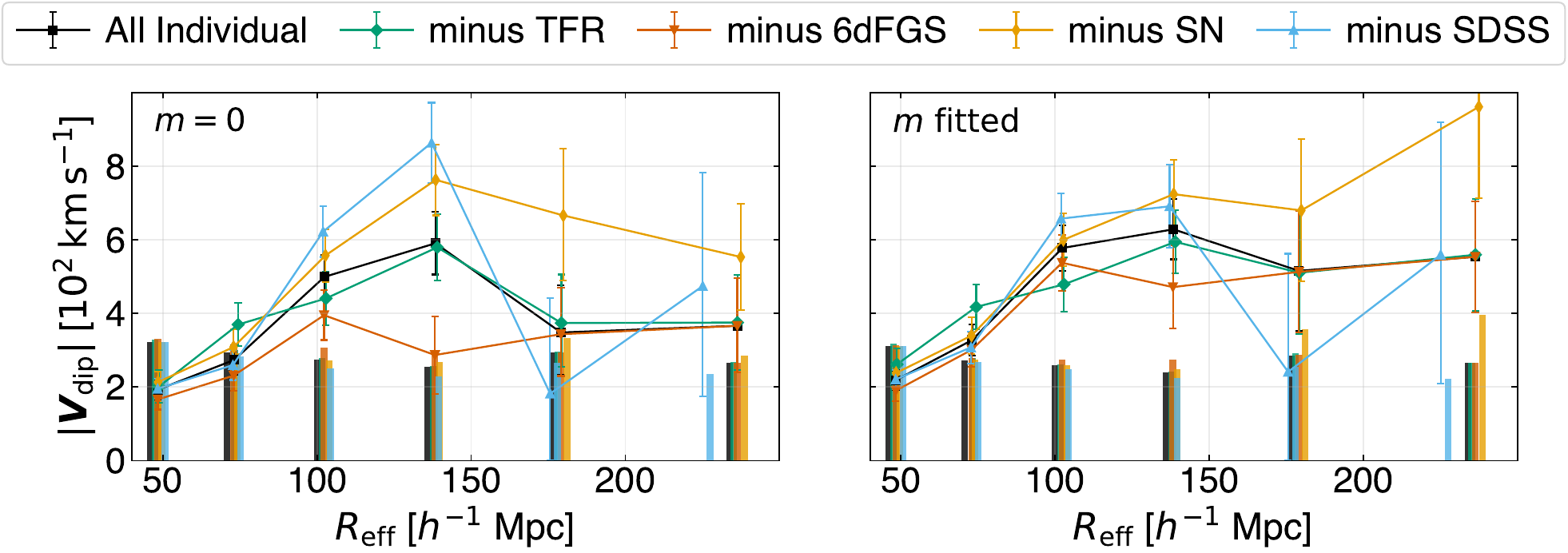}
\caption{Dipole amplitude after removing each major component catalog from
All Individual. The left panel fixes $m=0$ and the right panel fits $m$.
Capped error bars show the generally asymmetric observational
uncertainties; the asymmetry is small and difficult to see for some
points. Narrow vertical bars show the
catalog-specific \lcdm\ cosmic-variance ranges from zero to $V_{68}$, not
uncertainties centered on the measurements. Units are $10^2\,\kms$.}
\label{fig:leave_one_out_radial}
\end{figure*}

Table~\ref{tab:leave_one_out} gives the exact two-bin changes.
Figure~\ref{fig:leave_one_out_radial} shows the dipole-amplitude changes
over the full radial range for both the fixed- and free-monopole fits.
Other FP is omitted because its removal has negligible effect. As in
Figure~\ref{fig:amplitudes}, the observational errors on $|\Vdip|$ are
generally asymmetric, although the asymmetry is small and visually subtle
for several points.

\begin{table*}[t]
\centering
\scriptsize
\setlength{\tabcolsep}{5.0pt}
\caption{Local catalog-removal tests in the two least consistent bins}
\label{tab:leave_one_out}
\begin{tabular}{llrrrrrr}
\toprule
Removed & Bin & $N_{\rm obj}$ & $V_{\mathrm{SGX}}$ &
$\Delta V_{\mathrm{SGX}}$ & $|\Vdip|$ & $p_V$ & $p_{m+V}$ \\
& ($\hmpc$) & & ($\kms$) & ($\kms$) & ($\kms$) & & \\
\midrule
None     & 90--120  & 6,767 & $-545$ &   0 & 578 & 0.00208 & 0.00209 \\
SDSS     & 90--120  & 4,184 & $-612$ & $-68$ & 657 & 0.000314 & 0.000291 \\
6dFGS    & 90--120  & 4,670 & $-510$ & $+35$ & 537 & 0.0152 & 0.0239 \\
TFR      & 90--120  & 5,113 & $-461$ & $+84$ & 479 & 0.0281 & 0.0353 \\
SN       & 90--120  & 6,589 & $-551$ &  $-6$ & 600 & 0.00157 & 0.000829 \\
Other FP & 90--120  & 6,327 & $-550$ &  $-6$ & 585 & 0.00184 & 0.00193 \\
\addlinespace
None     & 120--160 & 8,481 & $-556$ &   0 & 628 & 0.000667 & 0.000111 \\
SDSS     & 120--160 & 3,875 & $-665$ & $-109$ & 691 & 0.000672 & $1.00\times10^{-5}$ \\
6dFGS    & 120--160 & 5,384 & $-411$ & $+145$ & 471 & 0.118 & 0.132 \\
TFR      & 120--160 & 7,980 & $-513$ &  $+43$ & 594 & 0.00196 & 0.000245 \\
SN       & 120--160 & 8,373 & $-633$ &  $-77$ & 724 & 0.000163 & $3.97\times10^{-5}$ \\
Other FP & 120--160 & 8,227 & $-544$ &  $+12$ & 630 & 0.000671 & 0.000126 \\
\bottomrule
\end{tabular}
\tablecomments{``None'' is the original All Individual result. Each other
row removes the named catalog and recomputes the fit and covariance.
$\Delta V_{\mathrm{SGX}}$ is relative to the original value in that bin.}
\end{table*}

\section{Discussion}
\label{sec:discussion}

\subsection{Robustness and limitations}
\label{sec:limitations}

The present calculation isolates the role of radial and angular windows,
but several effects remain outside the covariance model. Measurement
errors of different objects are treated as independent. We do not include
shared distance-calibration errors, repeated use of the same distance
measurement, or correlated errors between overlapping component catalogs.
The $100\,\kms$ floor is also independent between objects. It does not
replace a nonlinear correlated velocity model or a catalog-specific
small-scale dispersion.

The theoretical covariance has its own limits. The \lcdm\ calculation
uses the Planck 2018 cosmological parameters. It represents random \lcdm\
realizations and is not conditioned on the known density distribution in
our local Universe. The covariance is evaluated at $z=0$ and neglects the
mild evolution of the velocity field across the catalog.

Several observational inputs are inherited from CF4, including catalog
selection, Malmquist corrections, and distance calibration. A Malmquist
correction accounts for a selection-related bias in measured distances.
We do not refit the uncertainty in these corrections.

The linearized log-distance conversion is another approximation. The TFR
$120$--$160\,\hmpc$ result should be checked with an exact distance
likelihood. It should also be tested against the direct-velocity fit,
alternative velocity floors, modest shifts of the radial-bin boundaries,
and a Local Sheet frame analysis.

Table~\ref{tab:global_results} includes the full covariance between radial
bins for each catalog. A combined significance across catalogs would also
require the cross-catalog covariance and a predefined test statistic. The
current global rows must therefore not be combined as independent
measurements.

We inspected several catalogs, bins, and fitted components. The chance of
finding at least one small $p$-value increases when many tests are
examined. The quoted values have not been corrected for this effect. These
considerations motivate a selection-matched mock-catalog analysis that
includes calibration, overlap, and nonlinear velocity covariance.

\subsection{Interpretation}

The main interpretation is that CF4 contains a localized feature rather
than evidence for a uniformly anomalous flow over its full radial range.
The combined-bin tests include the covariance between shells, so this
conclusion does not result from treating the bins as independent. Instead,
the unusual motion is concentrated near $100$--$150\,\hmpc$ and is not
repeated with comparable strength at other radii. A localized feature may
still be physically important, but it is a different claim from a general
failure of the predicted large-scale velocity field.

The quantity measured here also differs from the bulk flows emphasized in
earlier CF4 studies. Our moments describe the objects in individual
shells through their actual catalog windows; they are not cumulative
top-hat flows, minimum-variance estimates of an ideal window, or
three-dimensional reconstructions. Different estimators weight the same
velocity field differently, so their amplitudes and significances need
not agree numerically. The shared qualitative result is enhanced coherent
motion on scales of roughly $100$--$160\,\hmpc$
\citep{Watkins2023CF4BulkFlow,Whitford2023CF4Estimators,
Courtois2025CF4pp}. The present analysis adds information about where that
feature enters the combined catalog and how strongly it depends on the
survey window.

The component catalogs do not identify one common anomalous mode. TFR is
most unusual in negative SGX across the common radial range, whereas the
local 6dFGS result arises from a tightly constrained combination of the
monopole and dipole. Other FP shows an isolated inner-bin feature, and SN
is more unusual in its monopole than in its dipole. These differences do
not necessarily imply catalog systematics: distinct angular and radial
windows can map the same cosmic velocity field into different fitted
moments. They do, however, prevent the small probabilities from being
interpreted as independent confirmations of one universal flow. All
Individual and its constituent catalogs are statistically related through
both shared measurements and cosmic covariance.

The SN component deserves separate attention because its distance reach
makes it potentially valuable despite its small size relative to the
galaxy samples. Its 1,093 objects, however, retain the historical sky
selection of the CF4 compilation and are highly nonuniform in declination.
The four declination bands $30^\circ<\delta<90^\circ$,
$0^\circ<\delta<30^\circ$, $-30^\circ<\delta<0^\circ$, and
$-90^\circ<\delta<-30^\circ$ cover equal areas of the sky, so an isotropic
sample would place 25 percent of its objects in each band. The actual
fractions are 27, 39, 27, and 7 percent, respectively. The sample is
therefore weighted toward the celestial north and the equatorial region,
limiting its ability to constrain low-order angular moments. Modern surveys
can obtain much larger SN samples over both hemispheres and to $z\sim0.1$.
ZTF DR2 already contains several thousand nearby SNe Ia, including a
volume-limited sample of about 1,000 through $z\simeq0.06$
\citep{Rigault2025ZTFDR2}. Samples of order 5,000 with more uniform angular
coverage should make SNe substantially more useful for studies of
large-scale velocity moments.

The leave-one-out tests measure influence, not attribution. Removing a
catalog changes the objects, weights, angular mask, and predicted cosmic
variance simultaneously. The strong response to removing 6dFGS therefore
does not assign a fixed fraction of the signal to that catalog or to a
particular structure. Its importance and the prominence of SGX are
qualitatively consistent with the Wiener-filter analysis of
\citet{Hoffman2024CF4Velocity}. The direction is also suggestive of the
Shapley Concentration, but establishing such an association would require
a velocity--density comparison rather than a velocity-only consistency
test. The direct SN--6dFGS comparison further shows that catalogs can pull
the combined fit in opposite directions while remaining statistically
consistent with each other.

The SDSS outer bins illustrate a separate limitation of sparse angular
coverage. There, the fitted monopole and SGY component are nearly
degenerate. Allowing the monopole to vary, fixing it to zero, or adding a
shared distance-scale uncertainty therefore changes the inferred dipole
and its significance. These alternatives do not merely change the error
bar; they define different estimators or covariance assumptions. The lack
of evidence for a quadrupole shows that an additional low-order angular
term is not required by these data, but it does not by itself establish
that the fitted outer dipole is a physical bulk motion.

Taken together, the results motivate targeted follow-up rather than a
claim of a robust challenge to \lcdm. The local and component-level tests
were identified after examining several correlated bins, catalogs, and
parameters. Their quoted significances describe the individual tests and
are not adjusted for the number of comparisons. The next step is a
selection-matched mock
analysis with predefined summary statistics, nonlinear velocity
covariance, correlated calibration and repeated-measurement errors, and an
exact distance likelihood for the largest residuals. Such a calculation
would determine whether the negative-SGX feature survives a unified
treatment of the observational and cosmological uncertainties.

\section{Conclusions}
\label{sec:conclusions}

We fitted shell-based monopole and dipole moments to CF4 catalogs using
their actual positions, weights, and catalog-specific \lcdm\ covariance.
The main conclusions are:

\begin{enumerate}
\item When all six All Individual bins through $300\,\hmpc$ are combined,
the dipole and joint deviations are $2.29\sigma$ and $2.36\sigma$,
indicating broad consistency with \lcdm. The strongest localized result is
the $120$--$160\,\hmpc$ bin, with $|\Vdip|=628\pm82\,\kms$ and
dipole and joint deviations of $3.40\sigma$ and $3.87\sigma$.

\item The localized excess is concentrated in negative SGX and depends
strongly on the catalog window. TFR and 6dFGS favor larger motions, while
SDSS and SN reduce the combined amplitude. Removing 6dFGS lowers the
upper-bin dipole deviation to $1.56\sigma$, but the standalone SN and
6dFGS dipoles remain mutually consistent. These tests measure catalog
influence, not an additive decomposition of the flow.

\item SDSS is consistent with \lcdm\ through $160\,\hmpc$; its outer-bin
result depends on whether the monopole is fixed and on the assumed shared
distance-scale covariance. The quoted significances use the baseline
covariance model and are not adjusted for the number of tests examined. A
robust cosmological assessment requires selection-matched mocks, correlated
calibration and nonlinear velocity covariance, and exact distance
likelihoods.
\end{enumerate}

\begin{acknowledgments}
This research has been supported by grant No.~893/22 from the Israel
Science Foundation and by a grant from the Asher Space Research Institute.
\end{acknowledgments}

\section{Data and code availability}

The CF4 input catalogs are available through the Extragalactic Distance
Database at \url{https://edd.ifa.hawaii.edu}. No new data products or
analysis code are released with this article.

\appendix
\renewcommand{\theHequation}{\Alph{section}.\arabic{equation}}

\section{From distance modulus to radial peculiar velocity}
\label{app:dm_velocity}

This appendix gives the sign convention, approximations, and uncertainty
calculation behind Equation~\eqref{eq:ulog}. It also clarifies the
difference between a distance-modulus offset and the fitted velocity
monopole.

\subsection{Distance ratio and the low-redshift limit}

Let $D_{L,i}$ be the luminosity distance inferred from a distance
indicator and let $D_{L,z,i}$ be the luminosity distance inferred from
the CMB-frame redshift in the adopted homogeneous cosmology. Their
distance-modulus difference is exactly
\begin{equation}
 \begin{aligned}
 \dmu_i
 &=5\log_{10}\left(\frac{D_{L,i}}{D_{L,z,i}}\right),\\
 \eta_i
 &\equiv
 \ln\left(\frac{D_{L,i}}{D_{L,z,i}}\right)
 =a\,\dmu_i,
 \qquad a=\frac{\ln 10}{5}.
 \end{aligned}
 \label{eq:dm_distance_ratio}
\end{equation}
Thus the measured distance is larger than the redshift distance when
$\dmu_i>0$.

The familiar low-redshift relation is
\begin{equation}
 cz_i\simeq H_0D_i+u_i,
 \label{eq:low_z_redshift_velocity}
\end{equation}
where positive $u_i$ is directed away from the observer. In this limit,
$D_{L,z,i}\simeq cz_i/H_0$ and
$D_{L,i}/D_{L,z,i}=\exp(\eta_i)$. A direct conversion would give
\begin{equation}
 u_{{\rm dir},i}
 \simeq
 cz_i\left[1-\exp(\eta_i)\right].
 \label{eq:low_z_direct_velocity}
\end{equation}
Expanding in the logarithmic distance ratio gives
\begin{equation}
 u_{{\rm dir},i}
 =-cz_i\eta_i+\mathcal O(cz_i\eta_i^2)
 =-cz_i a\,\dmu_i+\mathcal O(cz_i\dmu_i^2).
 \label{eq:low_z_log_velocity}
\end{equation}
A positive distance-modulus difference therefore corresponds to a
negative radial peculiar velocity. The object is farther away than its
redshift would indicate, so its peculiar motion is toward the observer
under our sign convention.

\subsection{First-order cosmological conversion}

The calculation does not replace the distance scale by $cz$. It instead
uses a redshift-dependent first-order coefficient. In the CMB-frame
calculation,
\begin{equation}
 D_{L,z,i}
 =\left(1+z_{{\rm hel},i}\right)
 D_C\left(z_{{\rm CMB},i}\right).
 \label{eq:appendix_redshift_luminosity_distance}
\end{equation}
The CMB-frame redshift determines the comoving position. The
heliocentric redshift supplies the observed photon-energy factor in the
luminosity distance \citep{Nusser2026DistanceVelocity}. We define
\begin{equation}
 \begin{aligned}
 \mathcal B_i
 &=1-
 \frac{H(z_i)D_C(z_i)}{c(1+z_i)},\\
 A_{U,i}
 &=\frac{H(z_i)}
 {(1+z_{{\rm hel},i})\mathcal B_i},\\
 K_i
 &=A_{U,i}D_{L,z,i},
 \qquad z_i=z_{{\rm CMB},i}.
 \end{aligned}
 \label{eq:appendix_velocity_scale}
\end{equation}
The factor $A_{U,i}$ converts a first-order luminosity-distance
perturbation to the comoving-coordinate radial velocity used here, while
$\mathcal B_i$ accounts for the first-order change in observed flux caused
by peculiar velocity, where ``first order'' refers to expansion in
$|u_i|/K_i$, not to a low-redshift or small-distance approximation. The
product $K_i$ has velocity units and tends to $cz_i$ at
low redshift. To first order, a radial comoving-coordinate velocity changes
the logarithmic luminosity distance by
\begin{equation}
 \delta\ln D_{L,i}=-\frac{u_i}{K_i}.
 \label{eq:velocity_distance_response}
\end{equation}
Combining Equations~\eqref{eq:dm_distance_ratio} and
\eqref{eq:velocity_distance_response} gives the estimator used in the
paper,
\begin{equation}
 u_{\ln,i}=-K_i a\,\dmu_i.
 \label{eq:appendix_log_velocity}
\end{equation}
It estimates the comoving-coordinate velocity. The corresponding
physical peculiar velocity is
$v_i=u_i/(1+z_i)$ to first order.

\subsection{Uncertainty, weighting, and the fitted monopole}

If the reported distance-modulus uncertainty is $e{\rm DM}_i$, then
\begin{equation}
 \sigma_{\eta,i}=a\,e{\rm DM}_i,
 \qquad
 \sigma_{u,i}^2
 =K_i^2\sigma_{\eta,i}^2+\sigma_{\rm floor}^2.
 \label{eq:appendix_velocity_error}
\end{equation}
The analysis treats the redshift and background cosmology as fixed when
forming this measurement variance. The velocity floor represents
independent small-scale scatter. It is added in velocity units and is
therefore equivalent to an effective distance-modulus variance
\begin{equation}
 \sigma_{\mu,{\rm eff},i}^2
 =e{\rm DM}_i^2
 +\left(\frac{\sigma_{\rm floor}}{aK_i}\right)^2.
 \label{eq:appendix_dm_error}
\end{equation}

The velocity-space model
\begin{equation}
 -K_i a\,\dmu_i
 =m+\widehat{\boldsymbol r}_i\mathbin{\cdot}\Vdip+\epsilon_{u,i}
 \label{eq:appendix_velocity_fit}
\end{equation}
can therefore be written equivalently as
\begin{equation}
 \dmu_i
 =-
 \frac{m+\widehat{\boldsymbol r}_i\mathbin{\cdot}\Vdip}
 {aK_i}
 +\epsilon_{\mu,i}.
 \label{eq:appendix_dm_fit}
\end{equation}
The two expressions give the same weighted least-squares solution when
the variances are transformed consistently. This shows why the fitted
$m$ is a constant velocity within a radial bin. It is not simply the
mean of $\dmu_i$ multiplied by one representative distance. The values
of $K_i$, the measurement weights, and the angular dipole term all enter
the estimate.

A common distance-modulus zero-point shift $\delta\mu_0$ produces
\begin{equation}
 \begin{aligned}
 \delta u_i
 &=-K_i a\,\delta\mu_0,\\
 \delta\mu_0
 &=5\log_{10}(1+\delta_D),
 &\delta u_i&\simeq-K_i\delta_D,\\
 C_{{\rm zp},ij}
 &=a^2 K_i K_j\sigma_{\mu_0}^2,
 \end{aligned}
 \label{eq:appendix_zero_point_velocity}
\end{equation}
where $\delta_D$ is a fractional distance-scale shift. If the common zero
point has variance $\sigma_{\mu_0}^2$, the last line is its object-level
velocity covariance. This covariance has rank one across the affected
objects before its contribution is mapped into the covariance of the
fitted moments. Its velocity scale grows with $K_i$. It is therefore not
identical to adding one constant
monopole to every radial bin. The dedicated SDSS zero-point test
includes this common shift as a correlated covariance term.

Finally, Equation~\eqref{eq:appendix_log_velocity} is linear in the
measured distance modulus. This preserves the approximately Gaussian
form of its reported errors. The direct transformation in
Equation~\eqref{eq:low_z_direct_velocity} is nonlinear and produces a
skewed velocity distribution even when the distance-modulus error is
Gaussian. For large $|\dmu|$, large fractional distance errors, or
strong outliers, an exact distance likelihood is preferable. Catalog
selection and Malmquist corrections are separate issues and are not
removed by either transformation.

\setcounter{equation}{0}
\section{Linear-theory velocity correlation functions}
\label{app:velocity_covariance}

This appendix gives the correlation functions used in
Equation~\eqref{eq:radial_covariance}. In linear theory at $z=0$, the
Fourier-space peculiar velocity is
\begin{equation}
 \boldsymbol v(\boldsymbol k)
 =iH_0f_0\,
 \frac{\boldsymbol k}{k^2}\,\delta(\boldsymbol k),
 \label{eq:linear_velocity_fourier}
\end{equation}
where $f_0=d\ln D/d\ln a$ is the present linear growth rate and $D(a)$
is the growth factor. At $z=0$, the physical and comoving-coordinate
velocity conventions coincide. We define the matter power spectrum by
\begin{equation}
 \left\langle
 \delta(\boldsymbol k)\delta^*(\boldsymbol k')
 \right\rangle
 =(2\pi)^3\delta_{\rm D}^{(3)}(\boldsymbol k-\boldsymbol k')P(k).
 \label{eq:power_spectrum_convention}
\end{equation}
For an isotropic $P(k)$, the
parallel and perpendicular velocity correlation functions are
\begin{align}
 \Psi_\parallel(s)
 &=\frac{H_0^2f_0^2}{2\pi^2}
 \int_0^\infty dk\,P(k)
 \left[j_0(ks)-\frac{2j_1(ks)}{ks}\right],
 \label{eq:psi_parallel_integral}\\
 \Psi_\perp(s)
 &=\frac{H_0^2f_0^2}{2\pi^2}
 \int_0^\infty dk\,P(k)
 \frac{j_1(ks)}{ks},
 \label{eq:psi_perp_integral}
\end{align}
where $j_0$ and $j_1$ are spherical Bessel functions. At zero
separation, both expressions approach
\begin{equation}
 \Psi_\parallel(0)=\Psi_\perp(0)
 =\frac{H_0^2f_0^2}{6\pi^2}
 \int_0^\infty dk\,P(k).
 \label{eq:psi_zero_separation}
\end{equation}

We use the $z=0$ linear matter power spectrum normalized to the adopted
Planck 2018 value of $\sigma_8$ and computed with the Eisenstein--Hu
transfer function \citep{PlanckCollaboration2018,EH98}. The numerical
calculation samples 2048 logarithmically spaced wavenumbers over
$10^{-4}\leq k/(h\,\mathrm{Mpc}^{-1})\leq10$ and tabulates the
correlation functions at 1024 separations out to $450\,\hmpc$. We use
the $z=0$ covariance throughout the low-redshift catalog and do not
include its mild redshift evolution.

\clearpage
\bibliographystyle{aasjournal}
\bibliography{Monopole_dipole_references}

\end{document}